\documentclass{emulateapj}    
\usepackage{epsfig}
\begin{document}

\title{Metallicity Distribution Functions of Four Local Group dwarf
  galaxies }

\author{Teresa L. Ross\altaffilmark{1}, 
  Jon Holtzman\altaffilmark{1},
  Abhijit Saha\altaffilmark{2},
  Barbara J. Anthony-Twarog\altaffilmark{3} }

\footnotetext[]{Based on observations made with the NASA/ESA Hubble Space
  Telescope, obtained at the Space Telescope Science Institute, which
  is operated by the Association of Universities for Research in
  Astronomy, Inc., under NASA contract NAS 5-26555. These observations
  are associated with program 12304. }
\altaffiltext{1}{Department of Astronomy, New Mexico State
  University, P.O. Box 30001, MSC 4500, Las Cruces, NM 88003-8001, 
  emails: rosst@nmsu.edu, holtz@nmsu.edu}
\altaffiltext{2}{NOAO, 950 Cherry Ave, Tucson, AZ 85726-6732}
\altaffiltext{3}{Department of Physics and Astronomy, University of
  Kansas, Lawrence, KS 66045-7582, USA, bjat@ku.edu}

\begin{abstract}
We present stellar metallicities in Leo I, Leo II, IC 1613, and
Phoenix dwarf galaxies derived from medium (F390M) and 
broad (F555W, F814W) band photometry using the Wide Field Camera 3
(WFC3) instrument aboard the Hubble Space Telescope. We measured
metallicity distribution functions (MDFs) in two ways, 1) matching
stars to isochrones in color-color diagrams, and 2) solving for the
best linear combination of synthetic populations to match the observed
color-color diagram.  The synthetic technique reduces the effect of
photometric scatter, and produces MDFs 30-50$\%$ narrower than the MDFs
produced from individually matched stars.  We fit the synthetic and
individual MDFs to analytical chemical evolution models (CEM) to quantify
the enrichment and the effect of gas flows within the galaxies. 
Additionally, we measure stellar metallicity gradients in Leo I and
II.  For IC 1613 and Phoenix our data do not have the radial extent
to confirm a metallicity gradient for either galaxy. 

We find the MDF of Leo I (dwarf spheroidal) to be very peaked with a
steep metal rich cutoff and an extended metal poor tail, while Leo II
(dwarf spheroidal), Phoenix (dwarf transition) and IC 1613 (dwarf
irregular) have wider, less peaked MDFs than Leo I. A simple
CEM is not the best fit for any of our galaxies, therefore we also fit
the `Best Accretion Model' of Lynden-Bell 1975. For Leo II, IC 1613
and Phoenix we find similar accretion parameters for the CEM, even though
they all have different effective yields, masses, star formation
histories and morphologies.  We suggest that the dynamical history of
a galaxy is reflected in the MDF, where broad MDFs are seen in
galaxies that have chemically evolved in relative isolation and
narrowly peaked MDFs are seen in galaxies that have experienced more
complicated dynamical interactions concurrent with their chemical evolution.

\end{abstract}


\section{Introduction}
Dwarf galaxies are important constituents of the universe because they
are both the most numerous type of galaxy and test beds for examining
galactic evolution on small scales.  Additionally, dwarf galaxies show
varying and extended periods of star formation (SF) over the age of the
universe \citep{2011ApJ...739....5W,2014ApJ...789..147W}. One method
of examining these populations is through metallicities, specifically
through metallicity distribution functions (MDF). Metallicity
measurements of large numbers of stars over the full range of
metallicities are crucial in studying the populations of dwarf
galaxies because they allow us to construct robust MDFs, examine any
structure indicating sub-components, and measure gradients across the
galaxy.  The star formation history (SFH), accretion, outflows via
supernova and stellar winds, and galaxy interactions that cause tidal
or ram-pressure stripping all play a part in shaping the MDF.

The shape of the MDF offers clues to the galaxy's evolution, which can
be characterized by fitting them to analytic chemical evolution
models.  The most basic analytic chemical evolution model is a simple  
closed box, where gas turns into stars, the stars evolve and return
enriched gas to the interstellar medium (ISM), which is again formed
into stars.  In this model no material enters or leaves the system.
However, the simple model is often an inaccurate description of the
chemical evolution in dwarf galaxies because the simple model
overpredicts the number of metal poor stars (i.e.~the G dwarf
problem).  Semi-analytic chemical evolution models
\citep{2010A&A...512A..85L,2014ApJ...785..102H} and hydrodynamic
simulations \citep{2006MNRAS.371..643M,2008MNRAS.386.2173M, 
2009A&A...501..189R}  utilize the galactic SFHs of dwarf galaxies, and
nucleosynthetic yields from type I and II supernova (SN), to predict
the gas inflow and outflow necessary to match abundance patterns from
multiple elements and the overall MDFs. In lieu of these complex
models the simple analytic chemical evolution models offer a concise
way to quantify and compare the MDFs of different galaxies, providing
insight into the enrichment history of galaxies, especially when only
modeling the overall MDF of the dwarf galaxy without information on
$\alpha$ abundances.

Recent works have examined the stellar metallicities found in dwarf
galaxies by using medium and  high resolution spectra on 8-10 meter
telescopes \citep{2003AJ....125..707T,2001ApJ...548..592S,
2003AJ....125..684S,2009AJ....137...62S,2011ApJ...727...78K,
2013ApJ...779..102K,2013A&A...549A..88S,2013A&A...554A...5K,
2014A&A...572A..88L,2014ApJ...785..102H}.  These works examine
variations in $\alpha$ elements, abundance patterns in the lowest
metallicity stars, abundance patterns in r and s process elements, and
the overall MDFs to better understand the processes that are important
in the chemical evolution; processes that include the SFHs, the IMF,
stellar nucleosynthesis, and the time-scales for the formation of
galaxies. One of the largest samples of stellar metallicities ($>$
3000 stars) comes from \citet{2011ApJ...727...78K,2013ApJ...779..102K}
who present spectroscopic MDFs for 15 dSph and 7 dIrs.  They also fit
their MDFs to analytical chemical evolution models, extended the
stellar-mass stellar-metallicity relation and determined that the MDF
shapes vary depending on the morphology such that dSph tend to have
narrower MDFs than dIrs.  

While individual spectra of stars in dwarf galaxies provide abundance
and kinematic information, they are still difficult to obtain. Spectra
of stars in more distant Local Group (LG) objects are limited to small
sample sizes by the long exposure times and large telescopes required
to make the observations.  While spectral targets are limited to the
few brightest stars, photometric metallicities probe deeper and thus
sample stars that are typically more representative of the metallicity
distribution.  Photometric metallicities, though not as accurate as
spectra, provide measurements for every star in the field, allowing
us to increase samples sizes by an order of a magnitude.  This allows
us to probe galaxies farther out in the LG, especially when using the
resolving power of the Hubble Space Telescope (HST).  We obtained HST
images in metallicity sensitive filters (F390M), and 2 wide band
filters (F555W and F814W) for four LG galaxies.  By measuring every
star in the field we can build up larger samples of stellar
metallicities in these dwarf galaxies.

In this work we present photometric metallicities of individual stars
in four LG dwarf galaxies: Leo I, Leo II, IC 1613 and Phoenix. We
chose galaxies that span different morphologies, masses, SFHs and
distances from the Milky Way (MW). Section 2 describes our two methods of
measuring the photometric MDFs, including a new synthetic color-color
diagram method, which is similar to the synthetic color magnitude
diagram (CMD) method of deriving SFHs. In section
3 we compare our metallicities to literature values, including recent
work with overlapping spectroscopic targets \citep{2011ApJ...727...78K,
2013ApJ...779..102K}. In Section 4 we fit the MDFs with chemical
evolution models to quantify the enrichment of the galaxy. In section
5 we present metallicity gradients from the central regions of each
galaxy. In section 6 we discuss implications of our results on
theories of galaxy evolution, and conclude in section 7.

\section{Observation}

\begin{deluxetable}{llcccc}[H]
\tablefontsize{\scriptsize}
\tablecaption{Observational data}
\tablehead{
 \multicolumn{1}{c}{Galaxy}&
 \multicolumn{1}{c}{Observation}&
 \multicolumn{3}{c}{ Total Exposure Time (sec)}\\
     \colhead{Name }&
     \colhead{dates }&
     \colhead{F390M }&
     \colhead{F555W }&
     \colhead{F814W }     }
\startdata
Leo I	&2011, Jun 02 $\&$ Mar 25	&21,024	&1,760	&1,500	\\
Leo II	&2012, Mar 24 $\&$ 25		&10,080	&1,760	&1,500\\
		&2013, Mar 30				&10,080	&---	
&---\\
IC 1613	&2011, Dec 16 $\&$ 20	&15,720	&1,200	&1,224\\
Phoenix	&2012, Jan 30 $\&$ 31	&16,320 &1,300	&1,340
\enddata
\label{obsdat}
\end{deluxetable}

The WFC3 observations were obtained between March, 2011 and March 2013
as part of HST proposal 12304 \citep{2008hst..prop11729H}. The dates
and exposure times of the observations are given in Table \ref{obsdat}.  
We used the reduced images from the STScI pipeline, which performs
bias, flat-field, and image distortion corrections. We additionally
used a charge transfer efficiency (CTE) correction module provided by
STScI \footnote{We used the alpha version (2013) for the CTE empirical
  pixel based corrections for WFC3/UVIS CTE located
  http://www.stsci.edu/hst/wfc3/ins\_performance/CTE/}.  
Magnitudes were measured using the point spread function fitting
photometry package DOLPHOT, which is a modified version of HSTphot
\citep{2000PASP..112.1383D}. The photometric catalog was culled to
reject objects based on goodness of fit and profile sharpness as
recommended by the DOLPHOT manual\footnote{The 2014 DOLPHOT 2.0 WFC3
  module  manual can be found at
  http://americano.dolphinsim.com/dolphot/}.

We adopt reddening and distance values reported in
\citet{2012AJ....144....4M} to calculate the absolute magnitudes. All
magnitudes are reported in the Vegamag system.  Table \ref{prop} lists
some basic observable quantities for these dwarf galaxies.

\begin{deluxetable*}{lllcclclllll}
\tabletypesize{\scriptsize}
\tablecaption{Dwarf galaxy properties}
\tablewidth{0pt}
\tablehead{
  \colhead{Galaxy}& 
  \colhead{Ra   } & 
  \colhead{Dec  } & 
  \colhead{D    } & 
  \colhead{M$_v$} & 
  \colhead{r$_h$} & 
  \colhead{M$_{\star}$} & 
  \colhead{M$_{Total}$} & 
  \colhead{$\sigma^{\star}$} & 
  \colhead{M$_{\rm{H I}}$} & 
  \colhead{$\sigma^{\rm{H I}}$} & 
  \colhead{type }  \\
    \colhead{ } & 
    \colhead{ } &   
    \colhead{ } &  
    \colhead{(kpc)} &
    \colhead{ } & 
    \colhead{('), (pc)} & 
    \colhead{($10^6 M_{\sun}$)}&
    \colhead{($10^7 M_{\sun}$)}&
    \colhead{(km s$^{-1}$)} &
    \colhead{($10^6 M_{\sun}$)}&
    \colhead{(km s$^{-1}$)} &
    \colhead{ }  }
\startdata 
IC 1613  	& 	
01 04 47.8 s &	
+02 07 04 	& 	
721 $\pm$ 5 &  	
-14.6 		&	
6.81, 1496  &	
100  		&	
79.5		&	
... 		&	
65  		&	
25  		&	
dIr      \\		
Phoenix  	&	
01 51 06.3 	&   
$-$44 26 41 &   
406 $\pm$ 13 &  
-10.1 		&   
3.75, 454   &   
0.77 		&   
3.3		 	&	
... 		&   
0.12 		&   
10  		&   
dTrans \\       
Leo I    	&	
10 08 28.1 s &  
+12 18 23 	&   
254 $\pm$ 17 &  
-11.9 		&   
3.40, 251   &   
5.5  		&   
2.2			&	
9.2 		&   
N/A  & N/A  &   
dSph     \\                                 
Leo II   	&	
11 13 28.8	&   
+22 09 06 	&   
233 $\pm$ 15 &  
-9.8  		&   
2.60, 176   &   
0.74 		&   
0.97		&	
6.6 		&   
N/A &N/A  	&   
dSph            
\enddata
\tablecomments{Data are taken from \citet{2009ARA&A..47..371T} and
  \citet{2012AJ....144....4M}. \citet{2012AJ....144....4M} reported the
  stellar mass, M$_{\star}$, assuming a stellar mass-to-light ratio of
  1. Virial masses reported from \citet{1998ARA&A..36..435M}. \label{prop}}
\end{deluxetable*}

\begin{figure*}[ht]
  \centering
  \epsfig{file=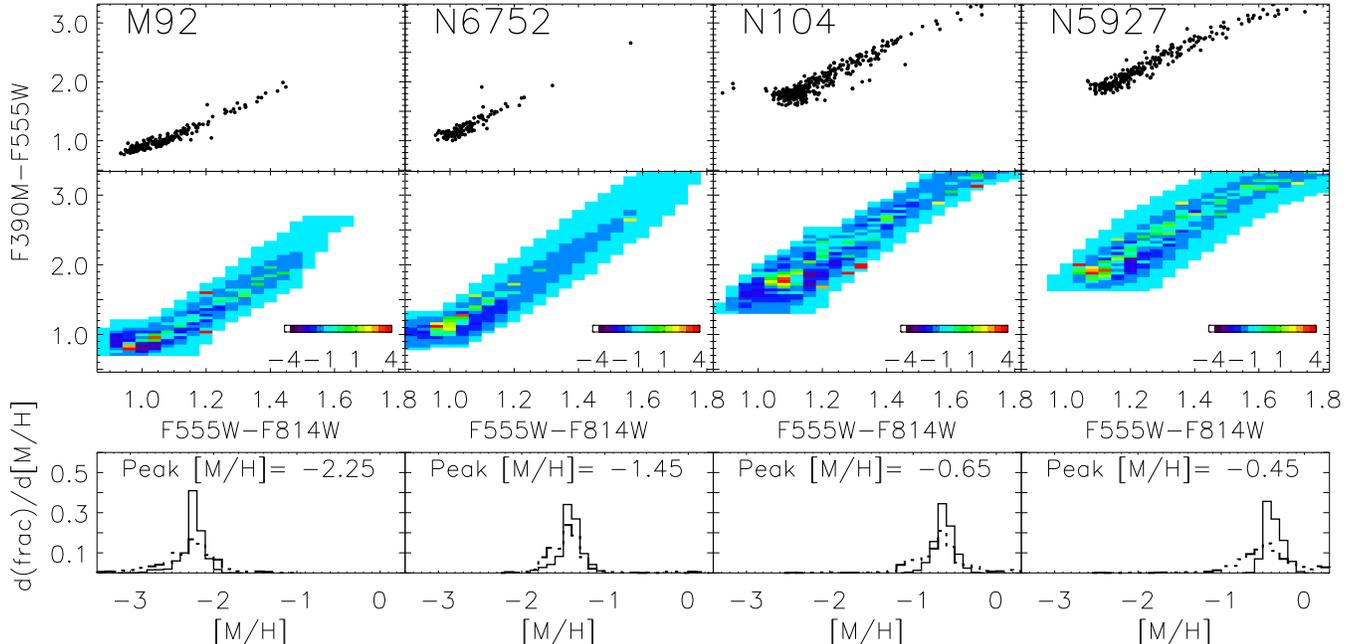,height=3.5in }
	\caption {The top panels show the color-color diagrams of giant
      branch stars in M92, NGC 6791, NGC 104 and NGC 5927. The middle
      panels show Hess diagrams of the residuals from the
      observations minus the best combination of synthetic populations
      normalized to the square root of the synthetic populations.  The
      bottom panels show the MDFs from the synthetic method in solid
      black as well as the MDF from individually matched stars, as
      dashed lines. \label{clusters}}
\end{figure*}

\section{Methods}
\subsection{Deriving Metallicity}

The general technique to measure photometric metallicities relates
color to metallicity.  The stellar properties that control observed
colors are metallicity, effective temperature and surface gravity.
However, effective temperature and surface gravity are dictated by the
mass, metallicity and the current evolutionary stage of a given star.
For populations of comparable age the color is directly related to the
metallicity.  However, an individual giant branch star can be redder
either because it has higher metallicity or because it is older,
leading to a color degeneracy between age and metallicity.  For a
mixed aged population, the younger giants are hotter and more massive
than older giants of the same metallicity so the color-metallicity
relation no longer holds.  
 
Using specifically designed HST WFC3 filters which separate the
effects of metallicity and temperature on color, we can break the
age-metallicity degeneracy.  The metallicity sensitivity of the
(F390M--F555W) color comes from the F390M medium band filter which
covers the Ca H $\&$ K spectral features, one of the strongest metal
absorption features in the visible spectrum. The temperature sensitive
color,  (F555W--F814W), uses two broad band filters which cover mostly
continuum.     

The color-color diagram (F390M--F555W, F555W--F814W), has been shown
to effectively separate the competing color changes due to metallicity
and temperature \citep{2014AJ....147....4R}, breaking the
age-metallicity degeneracy.  In the color-color diagram the color
change between a 12 and 4 Gyr isochrone of the same metallicity is on
order of a few hundredths of a magnitude, (demonstrated in Figure 5 of
Ross et al. 2014)\nocite{2014AJ....147....4R}.  Therefore, these two
colors can be used to measure individual stellar metallicities of
populations of mixed ages and abundances, like dwarf galaxies.

\subsection{Measuring Individual Stellar Metallicities}

We used only giant branch stars with errors less than 0.05 mag in all
filters. These error cuts resulted in samples of 3,449 stars, 444
stars, 896 and 578 stars for Leo I, Leo II, IC 1613 and Phoenix
respectively.

Stellar metallicities are assigned in the (F390M--F555W, F555W--F814W)
color-color diagram where each giant branch star is matched to the
closest Dartmouth isochrone \citep{2008ApJS..178...89D} in a grid
(spaced by 0.05 dex in [Fe/H] and assuming solar [$\alpha$/Fe]).  The
isochrones were empirically corrected to align with globular clusters
of known metallicities (M92, NGC 6752, NGC 104, NGC 5927 and NGC 6791
with [Fe/H] $=-2.30, -1.45,$ $-0.70, -0.40$ and $+0.40$
respectively). The empirical corrections were derived in
\citet{2014AJ....147....4R} using the same HST filters used in this
study. This method of deriving photometric stellar metallicities from
HST photometry was tested against the listed globular clusters and
shown to produce metallicities with errors of 0.15-0.3 dex, with
larger errors occurring at lower metallicities. See
\citet{2014AJ....147....4R} for the full details on the metallicity 
derivations from colors.   

Although each giant branch star is matched to an isochrone grid of
[Fe/H] assuming a solar abundance ratio, many stars are known to have
$\alpha$ enhancement, especially low metallicity ones.  The colors we
measure are a product of [Fe/H] and the $\alpha$ enhancement;
therefore, it is more accurate to think of our measurement as an
indicator of the total metallicity, [M/H], that represents the
intrinsic combination of [Fe/H] and [$\alpha$/Fe], even though we are
matching each star to an isochrone grid of [Fe/H]. Total metallicity,
[M/H], and the iron abundance, [Fe/H], are often used interchangeably,
because they are usually very similar; only in the cases of large
abundance variations (e.g $\alpha$ enhancements) do the two values
differ.  In \citet{2014AJ....147....4R} we quantified
the amount of color change expected from $\alpha$ enhancements and
provide a means of calculating the [Fe/H] if the $\alpha$ enhancement
is known.

\subsection{Synthetic Color-Color Diagram Method of deriving MDFs}
In addition to measuring metallicities of individual stars based on
color, we have also developed a synthetic color-color diagram
technique to derive a MDF from photometry.  The technique uses
isochrones and an initial mass function (IMF) to make a model
color-color diagram of the synthetic populations in order to reproduce
the observed color-color diagram.  

The technique is similar to the synthetic CMD method used to derive
SFHs (see Tolstoy et al.~2009). \nocite{2009ARA&A..47..371T}  In the
CMD the position and  number density of stars depend upon the IMF,
age, and metallicity. As pointed out by \citet{2002ASPC..274..450D},
the CMD of any complex stellar system will be a composite of all the
individual stellar populations that comprise the system. Therefore it
is possible to model a large number of simple stellar populations
(SSPs) in various combinations to reproduce the observed data,
including the observational errors.  The most likely combination of
populations of various ages and metallicities that best match the
observations will give the SFH. 

In the color-color diagram the position and number density of stars
depend upon metallicity and the IMF of a population, and is
independent of age, unlike a CMD.  The synthetic color-color diagram
method of deriving MDFs solves for the weights of a given linear
combination of populations that equals the observed color-color
diagram; the weights from the linear combination give the 
metallicity distribution of the population.  The advantage is that a
given population will occupy a locus where color and number density of
stars found at each location within the color-color diagram is
dictated by stellar evolution and the IMF \citep{2009ARA&A..47..371T}.

To create the synthetic color-color diagrams a Kroupa IMF
\citep{2001MNRAS.322..231K} and colors from the Dartmouth isochrones
are used to initially populate a grid of Hess diagrams of SSPs. The
isochrone colors were empirically corrected following the calibration
described in \citet{2014AJ....147....4R}. The grid spans a metallicity
range $-2.5 < \rm{[Fe/H]} <+0.5$ in steps of 0.05 dex. Each SSP Hess
diagram was blurred using the measured photometric errors from each
dwarf galaxy. The linear combination of SSPs that reproduces the
observed color-color diagram are calculated from the array of
synthetic Hess diagrams.  

The isochrone color spacing and the bin spacing of the Hess
diagrams (0.04 by 0.04 mag) require that the metallicity spacing (and
subsequent color change) be larger than the Hess bins. In some regions
of the color-color diagram isochrones spaced 0.05 dex of [Fe/H] apart
have a color change less than 0.04 mag. If the color change is less
than the bin size only one of the weights for the linear combination
of isochrones will be positive the rest will have negative and
nonphysical weights.  Additionally, in the synthetic color-color
diagram, when the synthetic SSPs are blurred by the photometric errors 
(up to $\sim$0.07 mag near the bottom of the giant branch), isochrones
closer than 0.15 dex produce nonphysical negative solutions.  To
account for the issues arising from the isochrone color spacing and
the bin spacing in the Hess diagrams, we combine the weights from
various metallicity combinations to ensure we sample all metallicities
equally.  We start with the full range of metallicities (-2.5 $<$
[Fe/H] $<$ 0.5, and spacing of 0.05 dex), calculate the linear
combination, eliminate the metallicities that have negative,
nonphysical solutions and recalculate the linear weights.  This leads
to uneven spacing in metallicity due to the smaller color separations
at low metallicity and larger color spacing at high metallicity.  We
repeat the calulation with different metallicity spacings (0.1, 0.15,
0.2, and 0.25 dex) and shifts of 0.05 dex in order to fill in the
metallicity spacings.  The weights from various metallicity
combinations are combine to ensure we sample all metallicities
equally.  

The MDF weights are checked by creating a synthetic color-color
diagram to compare to the observations.  We examine the residuals
between our observed Hess diagram and a synthetic Hess diagram to
ensure that $>99 \%$ of the residuals deviate by less than 3$\sigma$
of the overall residual within the Hess diagram. The residual Hess
diagram is created using the following equation:     

\begin{equation}
\rm{residual\ Hess\ diagram}= \frac{\rm{observed\ Hess -
    synthetic\ Hess} }{\sqrt{\rm{synthetic\ Hess}} }
\end{equation}

To test the synthetic color-color diagram method of deriving MDFs we
performed this analysis on globular clusters of known metallicity,
specifically M92, NGC 6791, NGC 104, and NGC 5927, with [Fe/H]$= -2.3,
-1.45, -0.70, -0.40$ respectively.  For these GCs we adopted all of
the parameters reported in \citet{2014AJ....147....4R}, where the
individual MDFs were derived.  Here after we will refer to the MDFs
derived using the synthetic color-color diagram method as `synthetic
MDFs', and the MDFs created from individually measured metallicities
as `individual MDFs'.  The synthetic MDFs recovered peak metallicities
that are all within 0.05 dex of the literature values.  Additionally
we find the synthetic distributions to be over two times as narrow as
the distributions found from fitting each star individually ($\sigma
=$0.15, 0.19, 0.24, and 0.19 for M92, NGC 6752, NGC 104 and NGC 5927,
respectively).  The narrower MDFs are expected because this method
accounts for the photometric errors, while the individually measured
metallicities do not.  

Figure \ref{clusters} shows the process and final results of the
synthetic color-color diagram method of deriving MDFs. The top panels
display the observed sequence of giant branch stars in the
color-color diagram. The middle panels show the Hess diagram of the
residuals as described in Equation (1).  The bottom panels show the
resulting MDFs from the synthetic method (solid line) as well as the
individual MDFs (dashed line).  The MDFs in the bottom panels
illustrate the utility of using the synthetic method over the
individual star method, wherein the synthetic MDFs are 30 - 50 $\%$
narrower than the individual MDFs. The synthetic method is statistical
in nature, therefore it does not provide individual metallicities for
each star, but rather gives the relative number of stars at each
metallicity for the population as a whole.

\section{Comparing Metallicities}

\subsection{Star by star metallicity comparison}

\begin{figure}[ht]
 \epsfig{file=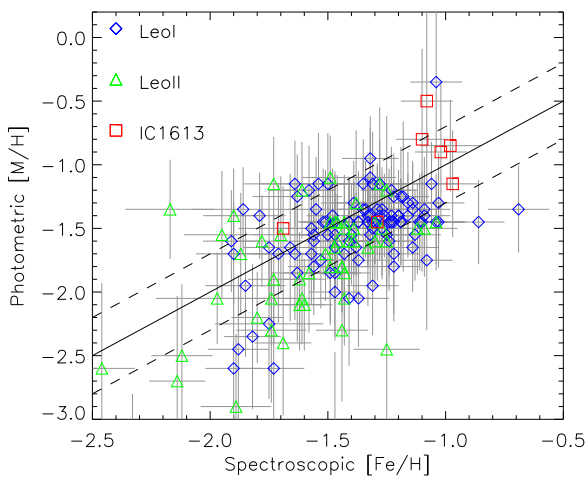, height=2.7in}
 \caption{The comparison of photometric and spectroscopic
   metallicities for 108 stars from Leo I, 54 from Leo II, and 8 from IC
   1613 shows that the photometric metallicities deviate to the metal
   poor side and the deviations are worse at lower metallicities.
   Spectroscopic metallicities measured by
   \citet{2011ApJ...727...78K,2013ApJ...779..102K}. \label{Z_comp}}  
\end{figure}

We directly compare the subset of our individually measured stellar
sample that overlap with spectroscopic measurements from
\citet{2011ApJ...727...78K,2013ApJ...779..102K} in Figure \ref{Z_comp}.  
\citet{2011ApJ...727...78K,2013ApJ...779..102K} used medium resolution
spectra and spectral synthesis of Fe I absorption lines to measure
metallicities for 814, 256, and 125 stars in Leo I, Leo II, and IC 1613,
respectively.  Due to the greater spatial extent of their sample we
were only able to compare a total of 170 stars;  108 from Leo I, 54
from Leo II, and 8 from IC 1613. The standard deviation of metallicity
differences is 0.38 dex, a wider spread (by 0.1 to 0.2  dex) than
found in the globular clusters from \citet{2014AJ....147....4R}.  
The larger spread between photometric and spectroscopic metallicity
can be partially attributed to the difference in photometric errors
between the dwarf galaxies and the globular clusters. The photometric
errors along the dwarf galaxies giant branch are 2 to 4 times larger
than equivalent absolute magnitudes found in the globular clusters,
which translates to a $\delta$[Fe/H] of $\sim$0.1 dex.

The larger spread might also be partially accounted for by $\alpha$
abundance variations. \citet{2014AJ....147....4R} find that variations in
$\alpha$ abundance cause color changes analogous to metallicity
changes of a few tenths of a dex (specifically $\Delta$[Fe/H]/$\Delta
\alpha$ $\sim$ $0.65-0.34$ across the metallicity range).  Without
knowing the intrinsic $\alpha$ abundance, and assuming it to
be solar, naturally our metallicity measurement will be lower than the
actual [Fe/H].  In general dwarf galaxies tend to show less $\alpha$
enhancements than globular clusters of similar metallicity.
\citet{2011ApJ...727...79K} find that the $\alpha$ abundance
distributions slowly evolve from large $\alpha$ enhancement at low
metallicity to roughly solar ratios at [Fe/H] close to a tenth
solar. Any $\alpha$ enhancement will cause an underestimate of the
metallicity using the photometric method.   

In addition to the greater difference between photometric and
spectroscopic metallicities, the photometric metallicities deviate
more towards lower metallicity. This is not unexpected as we assign
metallicities with an isochrone grid assuming no $\alpha$
enhancements. For any star that has an enhanced $\alpha$ abundance the
assigned metallicity from isochrones will be lower than the intrinsic
stellar metallicity because the color of a star of solar $\alpha$
abundance will have the same color as an $\alpha$ enhanced star with
lower [Fe/H].

The stars with photometric metallicities measured to be [M/H] $<-$2
tend to be the most discrepant, with differences $\geq $ 1 dex.
The main difference is due to the decreasing color change as a
function of decreasing metallicity,  which means that a small random 
variation in color for a bluer (metal poor) star will produce a larger
discrepancy in the reported metallicity than the same amount of
variation in a redder (metal rich) star.

\subsection{MDFs comparisons}

\begin{figure*}[ht]
  \centering
  \epsfig{file=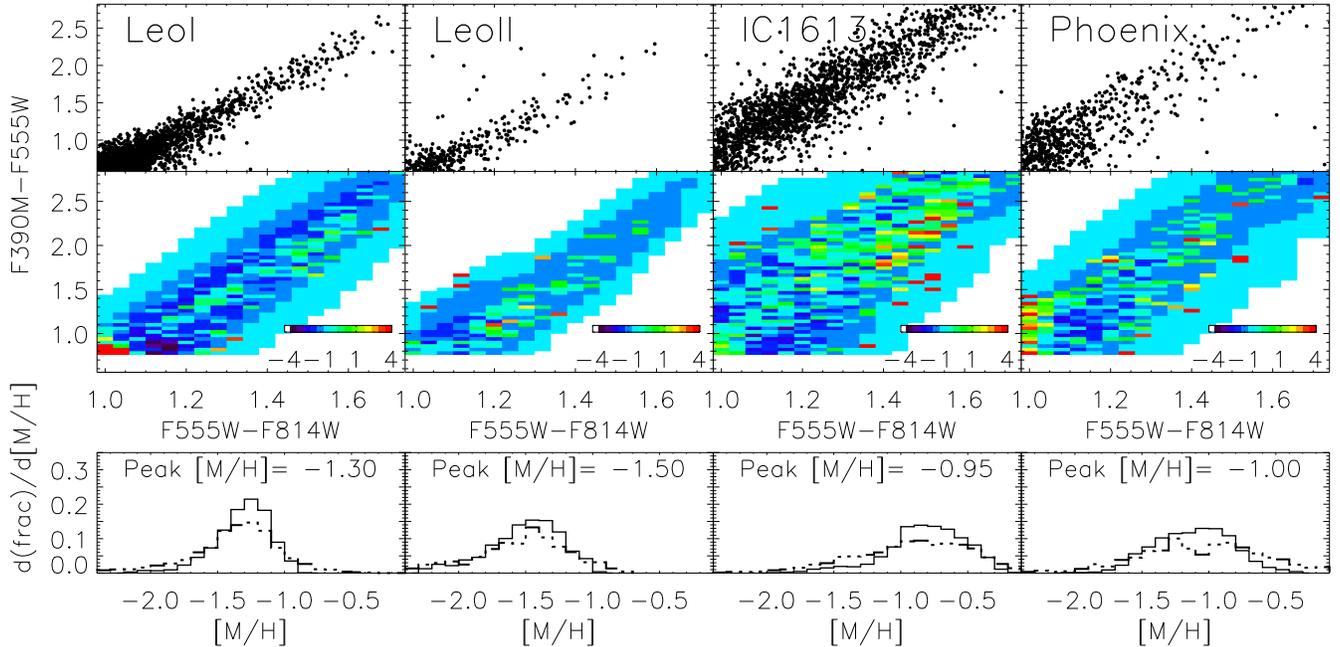,height=3.5in }
	\caption {The top panels show the color-color diagrams of giant
      branch stars in Leo I, Leo II, IC 1613 and Phoenix. The middle
      panels show Hess diagrams of the residuals from the
      observations minus the best combination of synthetic populations
      normalized to the square root of the synthetic populations.  The
      bottom panels show the MDFs from the synthetic method as the
      solid histogram, while the dashed lines are the MDFs derived
      from star individually matched to isochrones. \label{dwarfs}}
\end{figure*}

We used the synthetic color-color diagram technique to derive MDFs for
the four dwarf galaxies in addition to the individually measured
metallicities that were compiled to make MDFs.   Figure \ref{dwarfs},
following the same layout as Figure \ref{clusters}, shows the
results from the synthetic MDF derivation method. The top panels show
the color-color diagrams for the four dwarfs, the middle panels show
the residual Hess diagrams, and the bottom panels show the resulting
synthetic (solid line) and individual (dash-dotted line) MDFs. The
individual and synthetic MDFs both show similar shapes and peaks,
although the synthetic MDFs are narrower, as was noted in section 3.3
from the MDFs of globular clusters.

For the dwarf galaxies the average metallicities of the synthetic MDFs
are 0.1 to 0.2 dex more metal rich than the individual MDFs, although
the MDF peaks from the two methods are more closely aligned. 
Additionally, the globular cluster MDFs from the two methods are not
offset. This leads us to believe that the underlying population
distribution is causing the offset. For the dwarf galaxies the
difference in average metallicity from the two methods can be
attributed to the differences in the metal poor tails of the MDFs,
where the individual MDF has photometric errors that propigate into a
wider spread in metallicity.  Additionally, the fact that the color
change due to metallicity is smaller at low metallicities means that
for the same amount of color error there will be a larger spread in
metallicities at low values.  The large extent of the metal poor tail
in the individual MDF is the main driver in the offset of the average
metallicities for the 2 methods.

Despite the offset in average metallicity we believe the synthetic
method provides a more accurate MDF shape than the MDF produced by
individually measuring metallicities with isochrone matching.  The
synthetic method includes stellar sequence information from the
color-color diagram, and it systematically accounts for the
photometric errors, whereas the individual method relies only on the
photometry.

\begin{figure}[hT]
 \epsfig{file=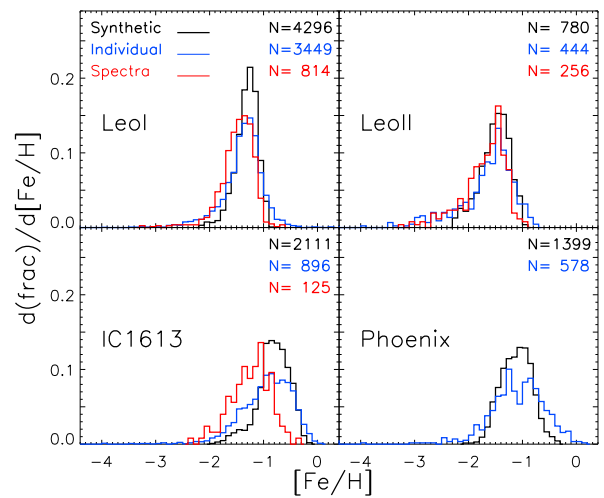, height=2.7in}
 \caption{The synthetic MDFs (black lines) are compared to the
   individual MDFs (blue lines).  For three of the galaxies
   spectroscopic MDFs (red lines) as measured by
   \citet{2011ApJ...727...78K} are also shown as a
   comparison.  The \citet{2011ApJ...727...78K} MDFs are true [Fe/H],
   as opposed to our measurements of [M/H], which are a combination of
   [Fe/H] and [$\alpha$/Fe].    \label{dwMDF_comp}}   
\end{figure}

In Figure \ref{dwMDF_comp}  we compare our two methods of deriving
MDFs to the spectroscopic MDFs available from
\citet{2011ApJ...727...78K,2013ApJ...779..102K}. Table \ref{MDFparam}
reports the average metallicities, widths, and number of stars
measured using the three different MDF derivation methods for each
galaxy.  The spectroscopic metallicities from 
\citet{2011ApJ...727...78K, 2013ApJ...779..102K} are more spatially
extended than our photometric data. Since we are  probing a smaller
area within each galaxy, it is expected that our MDFs will have some
differences compared to the spectroscopic MDFs. The results for each
dwarf galaxy are discussed in the following paragraphs.

For Leo I we find $<$[M/H]$_{individual}>$ $= -1.43$. The synthetic
MDF is narrow ($\sigma =0.16$), mostly Gaussian with an extension on
the metal poor end, and shows an abrupt cutoff on the metal rich side
of the distribution. The shapes, peaks and widths are consistent with
metallicities found in the literature. \citet{2011ApJ...727...78K}
found the median [Fe/H] $=-$1.42 ($\sigma$ = 0.33). 
\citet{2007MNRAS.378..318B} measured a MDF (for 101 RGB stars) peaked
at [Ca/H] $=-$1.34 ($\sigma$ = 0.21 dex), using the infrared
Ca-triplet method of measuring stellar metallicities. Using a
different Ca-triplet calibration \citet{2009A&A...500..735G} found
[Fe/H] $\simeq -1.37$ ($\sigma$ = 0.18 dex) using 54 stars.

For Leo II we find  $<$[M/H]$_{individual}>$ $= -1.77$  with the MDF
showing a more extended tail at low metallicities. The peak value and
overall MDF shape are consistent with those reported in the
literature. \citet{2011ApJ...727...78K} measured a MDF with a peak
[Fe/H] $= -1.71$ from the medium resolution spectra of 256 stars.  A
near IR photometric study of RGB stars in Leo II by
\citet{2008MNRAS.388.1185G} found a MDF peak, [M/H] $=-1.64$, when  
they account for the mean age (9 Gyr). \citet{2007MNRAS.378..318B}
measured stellar metallicities of 74 stars using the infrared
Ca-triplet of RGB stars, the MDF they measured peaked at [Ca/H]
$=-$1.65 ($\sigma$ = 0.17 dex). \citet{2007AJ....133..270K} also
measured spectroscopic metallicities using the Ca II infrared method
and found the mean metallicity of Leo II to be [Fe/H] $= -1.73$ from
52 stars.

For IC 1613  we find  $<$[M/H]$_{individual}>$ $=-0.99$, which is more
metal rich than the average spectroscopic metallicity measured by
\citet{2013ApJ...779..102K}, $<$[Fe/H]$>$ $= -1.19$.  Another average
metallicity measurement from \citet{2014ApJ...786...44S} finds
$<$[M/H]$>$ $= -1.3$ from a SFH study of a non-central field.  The
discrepancy between the higher average metallicity that we measure
compared to the values reported by \citet{2013ApJ...779..102K} and
\citet{2014ApJ...786...44S}, is partially accounted for by the fact
that we are looking at different regions of the galaxy, and thus a
younger population of stars. SFHs of IC 1613 show the star forming
regions move continuously inward over the lifetime of the galaxy
\citep{2007AJ....134.1124B}, where SF is associated with
higher average metallicity. Additionally, \citet{2003ApJ...596..253S} find
the stellar metallicities of IC 1613 to progress from [Fe/H] $=-1.3$ to
$-0.7$ over the age of the galaxy. Our field of view looks at the
inner $\sim$5 $\%$ of the galaxy, while the spectroscopic
metallicities are from stars spanning the whole galaxy and the 
SFH comes from a field 5.5' from the center. While the average
metallicity we find is more metal rich than most studies, we are more
metal poor than the [Fe/H] $= -0.67$ from the high resolution
study based upon 3 supergiants by \citep{2007AJ....134.2318T}. We
believe the average metallicity we measure is reasonable for the young
central region of IC 1613 that our field of view covers ($\sim 5\%$)
even though it is more metal poor than the average galactic
metallicity measured in other studies.

For Phoenix we find  $<$[M/H]$_{individual}>$ $=-1.17$. Upon first
examination this average metallicity is in conflict with metallicities
reported in the literature, although none of the literature values are
spectroscopically derived. Most of the reported metallicities are   
from SFH studies, like the one by  \citet{2000AJ....120.3060H} who found
$<$[Fe/H]$>$ $\sim -1.7$. The radial SFH study by
\citet{2013ApJ...778..103H} found the old population to have [Fe/H]
$\simeq -1.7$, and the younger populations to have [Fe/H] $\simeq
-1.2$.  Additional studies find [Fe/H] $=-1.9$, $-1.81 \pm 0.10$ 
\citep{1988PASP..100.1405O,1999A&A...345..747H} 
A study of variable RR Lyrae stars in Phoenix predicts $<$[Fe/H]$>$ $=
-1.55$ or $-1$.75 based on two different period metallicity relations 
\citep{2014ApJ...786..147O}.  According to the mass-metallicity
relationship \citep{2006ApJ...647..970L}, Phoenix, which is the same
stellar mass as Leo  II, should have a similar metallicity ([Fe/H] $\sim
-1.7$); yet, the average metallicity we measure is higher than the
galactic average metallicity reported in other studies.  

However, the field of view for our observations covers the innermost
region ($\sim 15 \%$) of Phoenix, where SF has occurred as recent as
100 Myr ago and higher metallicities are expected due to the
continuous star formation occurring in the central region.
Additionally, the radial SFH study by \citet{2013ApJ...778..103H} find
the central regions of Phoenix to  have a mean metallicity of $\simeq
-1.2$, which is in agreement with our results.  As with IC 1613, we
believe the higher average metallicity we measure reflects the greater
amount of enrichment that occurs in the central regions of dwarf galaxies.

\begin{deluxetable*}{lccccccccc}[H]
\tablefontsize{\scriptsize}
\tablecaption{MDF properties }
\tablehead{
 \multicolumn{1}{c}{Galaxy}&
 \multicolumn{3}{c}{Individual MDF}&
 \multicolumn{3}{c}{Synthetic MDF}&
 \multicolumn{3}{c}{Spectroscopic MDF}\\
   \colhead{Name }&
     \colhead{ $<$[M/H]$>$}&
     \colhead{$\sigma$}&
     \colhead{Num. of stars}  &   
     \colhead{ $<$[M/H]$>$}&
     \colhead{$\sigma$}&
     \colhead{Num. of stars} &    
     \colhead{ $<$[Fe/H]$>$}&
     \colhead{$\sigma$}&
     \colhead{Num. of stars}     }
 
\startdata
Leo I	&-1.43	&0.43 &3,449 &-1.34  &0.21 &4,296	&-1.45  &0.32   &814  \\
Leo II	&-1.77	&0.72	&444 &-1.54  &0.27   &780 	&-1.63  &0.40   &256  \\
IC 1613	&-0.99	&0.54	&896 &-0.84  &0.30 &2,111 	&-1.19  &0.37   &125  \\
Phoenix	&-1.17	&0.67 	&578 &-1.14  &0.28 &1,399	&---  	&---   	&---   
\enddata
\label{MDFparam}
\end{deluxetable*}

\section{Analytic models of Chemical Evolution}
The chemical enrichment within galaxies provides information on the
gas accretion, gas expulsion, the interaction history and star
formation history.  Matching analytic models of chemical evolution to
observations will constrain different evolutionary scenarios. It
stands to reason that dwarf galaxies could be represented by simple 
models due to their isolation and relatively less complicated
dynamics. Therefore we begin by fitting a simple model of chemical
evolution to our observed MDFs.

The simple model is essentially a leaky box, where gas is
allowed to leave the galaxy, but no accretion is occurring.
Additionally we assume $p$ is the effective yield, not the true
nucleosynthetic yield, where the effective yield is a product of both
the amount of metals created, and the amount blown out of
the galaxy through winds. The effective yield $p$ is in units of
Z$_{\sun}$.  The shape of the MDF for the simple model
follows the functional form as defined by \citet{1997nceg.book.....P}:

\begin{equation}
\frac{dN}{d{\rm[Fe/H]}} \propto  \left( 10^{\rm{[Fe/H]}} \right) 
{\rm exp} \left( - \frac{10^{\rm{[Fe/H]}}}{p} \right) 
\end{equation}

The simple model has a tendency to overproduce metal poor stars,
whether in dwarf galaxies or the solar neighborhood. We find that the
simple model is often unable to completely reproduce the MDF of dwarf
galaxies, in agreement with the findings of \citep{2011ApJ...727...78K, 
2013ApJ...779..102K}.  It has been theorized that the paucity of metal
poor stars can be explained by gas infall at later times
\citep{2008A&A...489..525P}.  Therefore we also tested an analytical
model which includes infalling gas. We fit the `Best Accretion Model'
defined by \citet{1975VA.....19..299L} which allows for the gas mass
to start small (or at zero), rise to a maximum, then approach zero
again as all the mass is accumulated in stars.  The accreted gas is
assumed to be metal free.  Lynden-Bell found a relation between the
gas mass, $g$,  stellar mass, $s$, and total final mass, $M$, that
permits an analytic solution to the differential metallicity
distribution (also see Pagel 1997)\nocite{1997nceg.book.....P},
defined as:

\begin{equation}
g(s) = \left( 1- \frac{s}{M} \right) \left(1+s-\frac{s}{M} \right)
\end{equation} 
where all quantities are in units of the initial gas mass. 
When $M=1$ the equation reduces to a closed box where $g=1-s$.  When
$M$ is larger than 1 it essentially becomes a measure of the total
amount of gas that has entered the system over the lifetime of the
galaxy so $M$ can be thought of as an accretion parameter.  However,
it should be noted that while Equation (3) produces the desired MDF
shapes, there is no physical rationale behind the actual form of the
equation.  Equation (3) simply assumes a quadratic relation between
gas mass and the stellar mass that peaks then decays as a function of
stellar mass. The MDF of the accretion
model, as defined by \citet{1997nceg.book.....P}, follows the form:

\begin{equation}
\frac{dN}{d\rm{[Fe/H]}} \propto \frac{10^{\rm{[Fe/H]}}}{p} 
\frac{1 +s(1-\frac{1}{M})}
{\left( 1-\frac{s}{M} \right)^{-1} -2\left( 1-\frac{1}{M} \right) 
 \frac{10^{\rm{[Fe/H]}}}{p} }
\end{equation}
where $s$ must be solved for numerically from the equation:

\begin{eqnarray}
{\rm [Fe/H]}(s)=&  \\ 
\log  \nonumber
 &\left\{  p \left( \frac{M}{1+s -\frac{s}{M}} 
	\right)^2  
  \left[ \ln \frac{1}{1-\frac{s}{M}} - \frac{s}{M}
 \left( 1-\frac{1}{M} \right) \right] \right\} 
\end{eqnarray}

For both the simple and the accretion model, the metallicity peak of
the model directly increases with $p$, as long as instantaneous mixing
of the ISM is assumed. The analytic form of the differential
metallicity distribution for these models requires the instantaneous
recycling and mixing approximations, where all SN yields are
immediately and uniformly returned to the ISM, respectively.  The
instantaneous recycling approximation does not reproduce the
characteristic patterns seen in $\alpha$ element abundances. 
However $\alpha$ abundances cannot be determined with photometric data
alone, therefore we could not constrain more sophisticated models that
account for time dependent recycling.  The analytic models used in
this work neglect some of the physics known to be important for
galactic chemical evolution, however, these models provide insight
into the difference in evolution between galaxies.

\subsection{MDF Truncation of the Metal Rich End}

For dSph, one leading hypothesis is that their chemical evolution was
interrupted by the removal of gas.  This follows from the
morphology-density relationship where most dwarf galaxies close to a
large primary are dSph, and devoid of gas \citep{2009ApJ...696..385G,
2014ApJ...795L...5S}. Environmental mechanisms such as tidal stirring
and ram pressure stripping have been invoked to transform close
satellites from a star forming, gas rich, dIr, into a quiescent, gas
poor dSph that are found close to a much larger primary
\citep{1983ApJ...266L..21L,2001ApJ...559..754M,2003AJ....125.1926G,
2011ApJ...726...98K}.  \citet{2006MNRAS.369.1021M} used simulations to
shows that gas rich dwarfs can be stripped of their gas over a few
pericentric passages as they travel through the hot halo of a larger
primary. Ram pressure stripping effectively cuts off all star
formation, stopping the chemical evolution prematurely, likely
producing a steep truncation on the metal rich end of the MDF. 

\citet{2013ApJ...779..102K} show that a sharp metal rich cutoff of
the MDF is primarily seen in dSph type galaxies, not dIrs.
Additionally, they added a ram pressure stripping term to the simple 
chemical evolution model to fit the metal rich cutoff, however, this
model was found to not have the best match to the dSphs in their
sample.  Since the accretion models fit all the dSph galaxies better than
the simple models we would like to incorporate a metallicity cutoff
into the accretion models, that would simulate ram pressure stripping.

A more basic approach to testing a metal rich cutoff is achieved by
changing the upper metallicity bounds to a more metal poor value when
integrating the equations of chemical evolution.  We ran the
simple and accretion chemical evolution models with an imposed metal
rich cutoff, truncating the CEM at a lower metallicity than the model
would naturally evolve to, in order to approximate the effects of ram
pressure stripping.  A sharp cutoff such as this is somewhat
nonphysical, since ram pressure stripping does not immediately remove
all gas, additionally the SF would not be immediately
cutoff at a particular metallicity across the galaxy.  However, this 
simplistic approach allows us to quickly and efficiently test our
models for signs of truncation.  

Besides the dSph galaxies, dIr are expected to have a steeper MDF on
the metal rich side than the chemical evolution models predict. 
The MDF of a star forming galaxy should not look like a completed
chemical evolution model, rather, the MDF should show a more abrupt
cutoff on the metal rich side because dIr are still forming stars and
enriching their ISM and by definition are not at the end of their
chemical evolution.  Fitting CEM with an imposed metal rich cutoff
will also simulate an incomplete chemical evolution, in addition to a
CEM that has been halted due to ram pressure stripping.

\subsection{Best fit models}
To find the best model parameters we perform 10$^3$ trials of the
simple model while varying the effective yield, $p$.  We calculate 
the least squares value between each model and MDF to determine the
best parameters for the individual and synthetic MDFs for each
galaxy. We also calculate the two sided confidence interval to find
the 1 $\sigma$ range of model parameters. For the Accretion Model we
preform 10$^4$ trials while varying the effective yield, $p$, and the
extra gas parameter, $M$. For the accretion + truncation models we run
10$^5$ trials varying the effective yield, $p$, the extra gas
parameter, $M$, and the cutoff metallicity. The best fit parameters
for each model and MDF are listed in Table \ref{cemparam}.

To fit the individual MDFs the three chemical evolution models have
been convolved with a Gaussian with a dispersion equal to the
dispersion of the difference between the photometric and spectroscopic 
metallicities ($\sigma$=0.38 dex) to account for the spreading of the
MDF from the photometric errors. The synthetic MDF method models 
the photometric errors, therefore we do not convolve the chemical
evolution models with any additional errors when fitting to the
synthetic MDFs.

\begin{figure}[ht]
 \epsfig{file=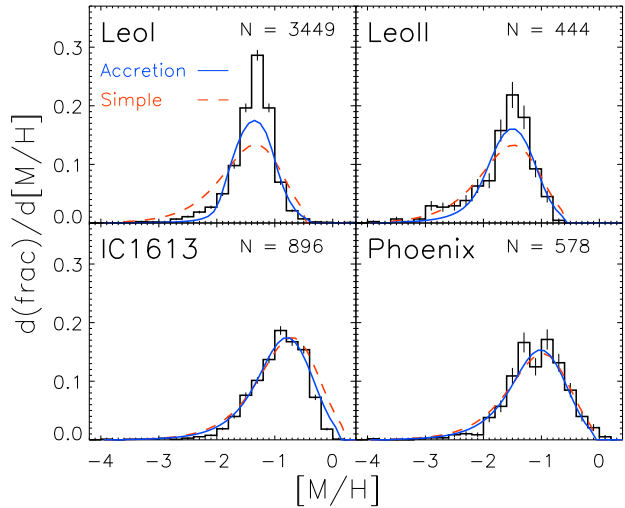, height=2.7in }
 \caption{The individual MDFs and Poisson uncertainties are shown with
   the best fit simple and accretion chemical evolution model. The
   number or stars in each MDF are listed for each galaxy.  \label{cem} } 
\end{figure}

\begin{figure}[ht]
 \epsfig{file=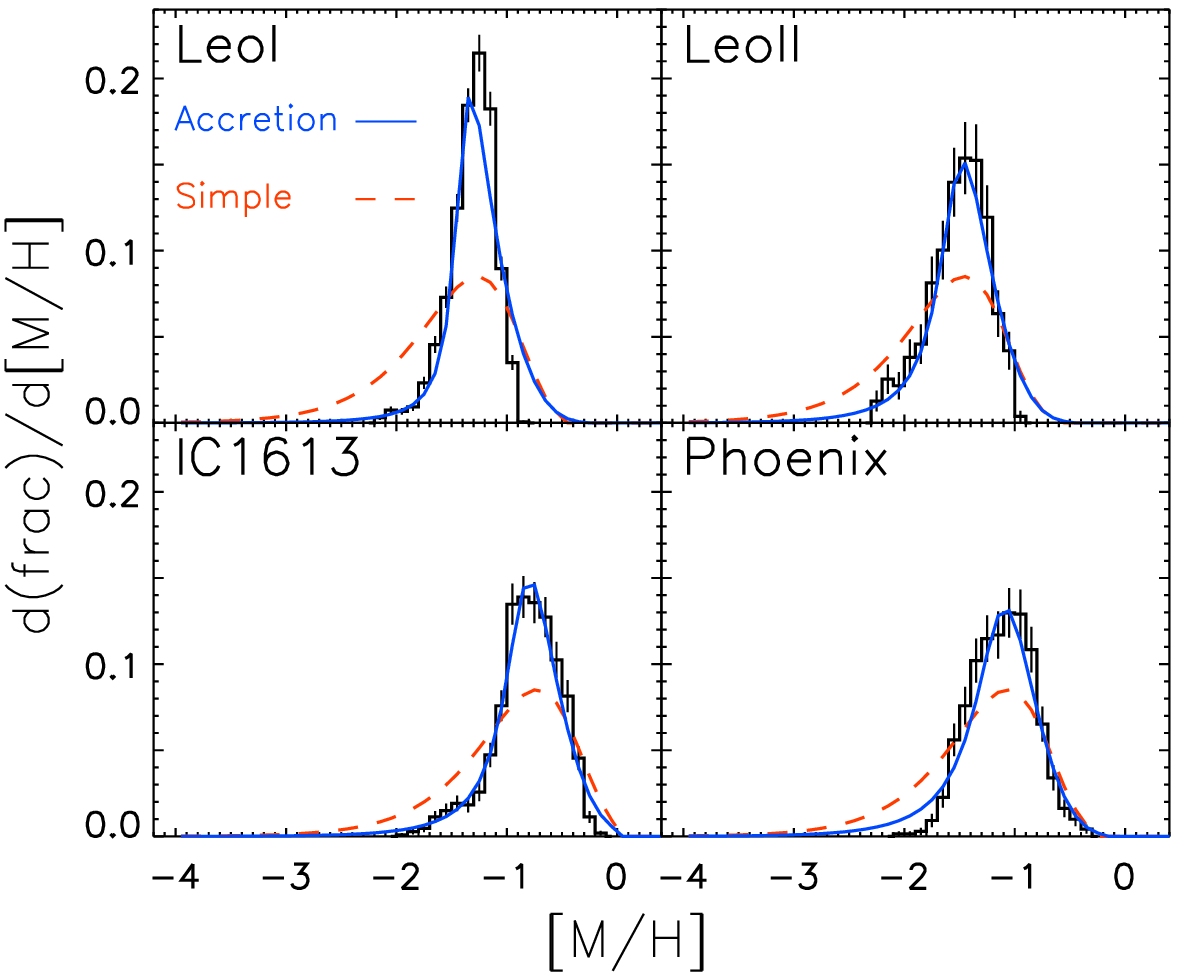, height=2.7in }
 \caption{the synthetic MDFs with Poisson uncertanties and the best fit
   simple and accretion chemical evolution models are shown.  The
   Poisson uncertainties are calulated from the total number of
   observed stars modeled by the synthetic color-color
   diagrams.  \label{cem_syn} }     
\end{figure}

\begin{figure}[ht]
 \epsfig{file=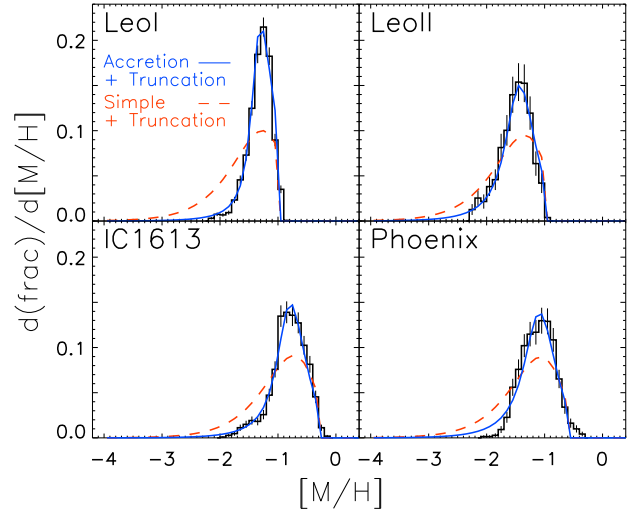, height=2.7in }
 \caption{ The synthetic MDFs with the Poisson uncertainties are shown
   with the best fit simple + truncation and accretion + truncation
   chemical evolution models. The Poisson uncertainties are calulated
   from the total number of observed stars modeled by the synthetic
   color-color diagrams.  \label{cem_syn_cut} }     
\end{figure}

The individual MDFs do not show an abrupt metal rich cutoff, mostly
due to the scatter in metallicity that comes from the photometric
technique as can be seen in Figure \ref{cem}. However, in the narrower
synthetic MDFs we do see truncation of the MDFs on the metal rich
side, indicative of ram pressure stripping. In the accretion model the
larger the M parameter the more the CEM takes on a Gaussian shape.
While this is good for fitting the narrow peak of the synthetic MDFs,
it does not account for the asymmetry of the extended metal poor tail
and the sharper truncation of the MDF on the metal rich side as shown
in Figure \ref{cem_syn}.  The accretion + truncation model allows for
a narrower MDF peak, and keeps the asymmetry by truncating the metal
rich side of the distributions, which produces a better model fit for
the synthetic MDFs in all four dwarf galaxies, as shown in Figure
\ref{cem_syn_cut}.

\subsubsection{Leo I} 
Leo I is one of the more distant MW companion dwarf spheroidal
galaxies (254 kpc, McConnachie 2012); \nocite{2012AJ....144....4M}  
additionally it is receding quickly from the MW ($V_{rad}$=167.9 km
s$^{-1}$, $V_{tan}=$101 km s$^{-1}$, Sohn et al.~2013),
\nocite{2013ApJ...768..139S} making it one of the 
more unusual dSph galaxies. Since its discovery in the 1950's
\citep{1950PASP...62..118H} Leo I has been studied by many groups to
better understand the SFH, dynamic properties,
metallicities and chemical enrichment history.  Here we compare our
chemical evolution models to results from previous studies. 

We find the accretion + truncation model is the best fit
model for the synthetic MDF.  The synthetic MDFs shows a narrow peak at
[M/H] $\simeq -1.35$, a slight metal poor tail, and an abrupt cutoff
on the metal rich side of the distribution. The simple model
underpredicts the number of stars in the peak, and overpredicts the
amount of metal poor stars.  The accretion model fits the MDF better
than the simple model, however this model does not fit the truncation
on the metal rich side of the MDF, and subsequently overpredicts the
number of metal rich stars. We find that the narrowness of the
synthetic MDF, and the metal rich truncation is best fit by the
accretion + truncation models, with a large accretion
parameter of $M=6.60^{+2.2}_{-2.8}$, yield of $p=0.07^{+0.005}_{-0.005}$ and
a cutoff metallicity of [Fe/H] $=-$1.0. 

\citet{2011ApJ...727...78K,2013ApJ...779..102K} also modeled the MDF
of Leo I, with a simple, pre-enriched and accretion model. They found
the accretion model was the best fit, with $M=7.9$ and $p=0.043$,
which is within errors but smaller than our accretion model parameter
of $M=9.8^{+3.1}_{-3.0}$ and the yield we derive, $p=0.068$, is higher
than the one they derive, but our MDF has a higher average metallicity
(see discussion in Section 4.2).

\citet{2010A&A...512A..85L} created detailed chemical evolution
models for Leo I and II, that use nucleosynthesis for type I and II
SN, an exponentially decreasing accretion parameter, the SFH, and
$\alpha$ abundances and the MDF as inputs to solve for the outflowing
wind, and star formation efficiency.  The detailed chemical evolution
model that best matches the Leo I abundances predicts that the galaxy
started with a low star formation efficiency (0.6 Gyr$^{-1}$) for the
first 5 Gyr, then had another 7 Gyr episode of SF starting
at 9 Gyr with substantial galactic winds throughout
\citep{2010A&A...512A..85L}.  
The evolution model of \citet{2010A&A...512A..85L} agrees with the 
color magnitude diagram (CMD) based SFHs showing that Leo I had slow
ancient SF, with an increase in the SFR around 7 - 8 Gyr ago, and
another increase to its highest rate 2-3 Gyr ago, after which the
galaxy stopped forming stars 1 Gyr ago \citep{1999AJ....118.2245G,
1999AJ....117.2199C,2000MNRAS.317..831H,2002ASPC..274..450D,
2014ApJ...789..147W}.

\subsubsection{Leo II} 
Leo II is also one of the more distant MW companion dwarf
spheroidal galaxies (233 kpc, McConnachie 2012). \nocite{2012AJ....144....4M} 
Unlike Leo I, the proper motion study of Leo II reveals the
galactocentric velocity is mostly tangential ($v_{tan}$ = 265.2 $km
s^{-1}$), with only 21.5 $km s^{-1}$ in the  radial direction.
Subsequent dynamic modeling of the large velocity and large distance
of Leo II suggest that Leo II is most likely passing into the MW halo
for the first time \citep{2011ApJ...741..100L}.   

The synthetic MDF of Leo II is best fit by the accretion + truncation
model with an accretion parameter of $M=4.0^{+2.0}_{-2.4}$, yield of
$p=0.038^{+0.05}_{-0.008}$ and a metallicity cutoff of [Fe/H] $=-$1.0. 
The simple model underpredicts the number of stars found in the MDF
peak, and overpredicts the number stars in the metal poor tail. The
accretion model fits the peak and metal poor tail, but over predicts
the number of stars on the metal rich end of the distribution,
explaining why the best fit model includes the metallicity truncation
term.  

Our chemical evolution results are in agreement with
\citet{2011ApJ...727...78K,2013ApJ...779..102K} who found the
accretion model to be the best fit for Leo II, with an accretion
parameter of $M=3.1^{+0.6}_{-0.5}$ and yield of $p=0.028$.  For our
accretion model we measured $M=4.1^{+3.0}_{-3.1}$ and $p=0.038$.
\citet{2010A&A...512A..85L} modeled the detailed chemical evolution
of Leo II based on the overall MDF and the $\alpha$ element abundance
ratios measured by \citet{2009AJ....137...62S}.  Their best fit model
predicts one long (7 Gyr) episode of SF starting at 14
Gyr, with a lower SF efficiency than seen in Leo I (0.3
Gyr$^{-1}$) with very high wind efficiency throughout.  The chemical
evolution of \citet{2010A&A...512A..85L}  is in good agreement with
the SFH derived from the wide field photometric survey of
\citet{2007AJ....134..835K}. They found that Leo II evolved with a low
SF rate up to 8 Gyr ago when SF stopped in the
outer regions, from 8 to 4 Gyr ago the  central star forming region
continually shrank until SF essentially stopped.

\begin{deluxetable*}{lccccccll}[H]
\tablecaption{Chemical evolution model parameters}
\tablehead{
 \multicolumn{1}{c}{Galaxy}&
 \multicolumn{1}{c}{MDF type} &
 \multicolumn{1}{c}{Simple Model}&
 \multicolumn{2}{c}{Accretion Model}&
 \multicolumn{3}{c}{Accretion + truncation}&
 \multicolumn{1}{c}{Best Model}\\
     \colhead{Name }&
     \colhead{}	  &
     \colhead{$p$}&
     \colhead{$p$}&
     \colhead{$M$}&
     \colhead{$p$}&
     \colhead{$M$}&
     \colhead{[Fe/H] cutoff}&
     \colhead{}     }
\startdata
Leo I	&Indiv
	&0.054 $^{+0.008}_{-0.002}$ 				   
	&0.054 $^{+0.016}_{-0.026}$  &8.2$^{+1.9}_{-2.4}$	   
	&0.050 $^{+0.020}_{-0.026}$  &7.7$^{+1.9}_{-2.2}$ &-0.10 
	&A+T	\\
Leo II	&Indiv
	&0.038 $^{+0.008}_{-0.006}$ 				
	&0.038 $^{+0.010}_{-0.008}$ 	&3.0$^{+1.3}_{-1.9}$	
	&0.038 $^{+0.040}_{-0.008}$	&3.0$^{+1.3}_{-2.0}$ &-0.40 
	&S	\\
IC 1613	&Indiv
	&0.228 $^{+0.041}_{-0.045}$ 				 
	&0.184 $^{+0.034}_{-0.032}$ 	&1.5$^{+1.2}_{-0.5}$	
	&0.182 $^{+0.036}_{-0.032}$	&1.7$^{+1.5}_{-0.7}$ &-0.30 
	&A+T	\\
Phoenix		&Indiv
	&0.118 $^{+0.015}_{-0.012}$  				
	&0.112 $^{+0.020}_{-0.014}$ 	&1.5$^{+0.7}_{-0.5}$	
	&0.112 $^{+0.027}_{-0.018}$	&1.5$^{+1.1}_{-0.5}$	&-0.20 
	&A\\
\hline  \\
Leo I	&Synth 
	&0.054 $^{+0.013}_{-0.010}$  			      
	&0.068 $^{+0.012}_{-0.012}$ 	&9.8$^{+2.1}_{-2.0}$ 
	&0.070 $^{+0.005}_{-0.005}$ 	&6.6$^{+2.2}_{-2.8}$ &-1.0 
	&A+T	\\
Leo II	&Synth	
	&0.036 $^{+0.044}_{-0.018}$   				
	&0.038 $^{+0.022}_{-0.016}$	&4.1$^{+2.4}_{-2.1}$ 
	&0.038 $^{+0.050}_{-0.008}$	&4.0$^{+2.0}_{-2.4}$	&-1.0 
	&A+T	\\
IC 1613		&Synth
	&0.182 $^{+0.026}_{-0.082}$   			
	&0.196 $^{+0.012}_{-0.070}$	&4.1$^{+2.0}_{-2.1}$ 
	&0.200 $^{+0.010}_{-0.036}$	&3.7$^{+2.0}_{-1.9}$ &-0.40 
	&A+T\\
Phoenix		&Synth
	&0.086 $^{+0.050}_{-0.042}$   				
	&0.092 $^{+0.058}_{-0.038}$ 	&3.4$^{+2.0}_{-2.4}$ 
	&0.094 $^{+0.042}_{-0.018}$	&3.3$^{+2.6}_{-1.6}$	&-0.60 
	&A+T

\enddata
\label{cemparam}
\end{deluxetable*}

\subsubsection{IC 1613:}

As one of the nearest gas rich dwarf irregular galaxies at a distance
of 721 kpc there have been numerous studies of the gas and stars in IC
1613 \citep{2012AJ....144....4M}.  IC 1613 is similarly distant from
M31 (517 kpc) however it is not considered to be a satellite of either
the MW or M31 \citep{2012AJ....144....4M},  but without data on the
proper motion the membership of IC 1613 to either is not definitive.
Various H {\footnotesize I} studies have measured the total H
{\footnotesize I} mass as 3 - 6 x$10^7 M_{\sun}$
\citep{1989AJ.....98.1274L,2006A&A...448..123S}. Additionally, the
studies show complicated morphology including numerous H
{\footnotesize I} arcs, holes and shells produced from the ongoing
SF within IC 1613 as evidenced by the presence of OB
associations and H {\footnotesize II} regions  \citep{2010A&A...523A..23G}. 

For IC 1613 we find the accretion + truncation model with
an accretion parameter of $M=3.7^{+2.0}_{-1.9}$, a yield of
$p=0.20^{+0.008}_{-0.036}$ and a cut off metallicity of [Fe/H] $
=-$0.40 is the best fit to the synthetic MDF. IC 1613's MDF shows a
broad peak at [M/H] $=-1.00$, an asymmetric Gaussian shape with an
extended metal poor side of the distribution.  The simple model
underpredicts the number of stars in the peak, overpredicts the metal
poor tail, and slightly overpredicts the number of stars on the metal
rich side of the MDF.  The difference between the accretion model and
the accretion + truncation model is small, and almost
entirely due to the metal rich side of the MDF. Since IC 1613 is an
isolated galaxy that has a substantial gas mass and shows no sign of
interaction, ram pressure stripping cannot be the process truncating
the MDF. Instead, the steeper metal rich side of the MDF reflects the
continuing chemical enrichment of IC 1613.  

Comparing our work with \citet{2013ApJ...779..102K} we find that our
accretion model with $M=3.7^{+3.0}_{-1.9}$ and $p=0.20^{+0.01}_{-0.036}$ 
has a similar accretion parameter but a higher effective yield
compared to their $M=4.3^{+1.5}_{-1.1}$ and yield of $p=0.075$.  The
higher effective yield we measure is potentially due to the different
regions of the galaxy sampled.  We cover a small ($\sim 5\%$) central
field while the metallicities from \citet{2013ApJ...779..102K} cover
the entire galaxy. Even though our accretion models find similar
parameters, \citet{2013ApJ...779..102K} find that a pre-enriched model
has a more likely fit than the accretion model. The pre-enriched model
is a generalization of a closed box model that starts with some
initial metallicity and does not accrete additional gas over its lifetime.

\subsubsection{ Phoenix} 
Phoenix is a transition type galaxy at a distance of 415 kpc and an
absolute V magnitude of -9.9 \citep{2012AJ....144....4M}.  There is no
H {\footnotesize I} emission detected within Phoenix itself, however,
H {\footnotesize I} gas has been detected at 4.5' to 9' from the
center of the galaxy \citep{2007ApJ...659..331Y} indicating recent gas 
expulsion. Young et al.~also state that the expelled gas is likely
linked with the most recent episode of SF ($\sim$100
Myr ago; Bianchi et al.~2012). \nocite{2012AJ....143...74B} 
 
We find the accretion + truncation model to be the best
fit model for the MDF, with an accretion parameter of $M=3.3^{+2.6}_{-1.6}$,
a yield of $p=0.094^{+0.042}_{-0.018}$, and a cut off metallicity of
[Fe/H] $=-0.6$.  The simple model underpredicts the number of stars in
the MDF peak and overpredicts the amount of metal poor stars in the
tail of the distribution.  The difference between the accretion model
and the accretion + truncation is small; both models 
slightly overpredict the number of metal poor stars while on the metal
rich side of the MDF the accretion + truncation model has
a slightly better agreement.

Ground-based photometry of Phoenix by \citet{1999AJ....118..862M}
noted the presence of a structurally distinct inner population with
mostly young stars in the east–west direction, while isophots of the
older component are aligned in the north-south direction.
Another ground based wide-field (26 x 26 arcmin$^2$) photometric study
shows the young ($<$ 1 Gyr) inner population to have a disk like
distribution \citep{2012MNRAS.424.1113B}. While the accretion +
truncation model is found to be the best fit, evidence for ram
pressure stripping is not seen in any of the population studies, nor
would it be expected in an isolated galaxy.  

In a study of the radial SFH properties of dwarf galaxies
\citet{2013ApJ...778..103H} additionally modeled the chemical
evolution for Phoenix.  Their best fit model predicts an initial
episode of SF lasting 3-4 Gyr with very little or no metal
enrichment that accounts for 75$\%$ of the total cumulative stellar mass.
Additionally, the SF episode is longest in the central
regions with shorter SF episodes in the outer regions. To
account for the minimal amount of enrichment ([Fe/H] $=-1.67$) in this
first phase their model predicts a high SFR that would expel most of
the metals formed in this episode before they can mix with the
interstellar medium.  The second phase of SF produces most
of the metal enrichment, and mainly occurs in the central regions
where the gas density is still sufficiently 
high to support continued SF, they find the metallicity of this burst
to be [Fe/H] $=-1.08$.  We find the MDF peak to be [M/H] $= -1.17$,
which is consistent with the enrichment values \citet{2013ApJ...778..103H}
found in the central regions of Phoenix from their radial SFH study. 
However, much like the initial metallicity \citet{2013ApJ...778..103H}
found, other photometric studies measured the overall metallicity
in Phoenix as $<$[Fe/H]$>$ $\sim$  -1.7, \citep{2000AJ....120.3060H},
[Fe/H] $=-1.9$, \citep{1988PASP..100.1405O}, $<$[Fe/H]$> =-1.81
\pm0.10$ \citep{1999A&A...345..747H}.

\section{Metallicity Gradients}

Metallicity gradients found in stellar populations offer clues about
past SF, as well as interaction history since mergers are
thought to remove gradients. Many dwarf galaxies have
detectable metallicity gradients  (e.g.~Fornax, Cetus, Carina,
Sculptor, Sextans, Tucana, Leo I, Leo II, and Draco, Andromeda I-III,
V and VI. From Harbeck et al.~2001; Tolstoy et al.~2004; Battaglia et
al 2006; 2011, 2012; Monelli et al 2012; Kirby et al.~2011), 
\nocite{2001AJ....122.3092H}  \nocite{2004ApJ...617L.119T}  
\nocite{2006A&A...459..423B,2011MNRAS.411.1013B,2012ApJ...761L..31B}
 \nocite{2012MNRAS.422...89M} while others like Ursa Minor and Canes
 Venatici do not show  measurable gradients (Kirby et al.~2011).
 \nocite{2011ApJ...727...78K} 

Metallicity gradients can be produced by increased SF (and
thus enrichment) in the central regions due to increased gas supply 
there. The gas density increases with the gravitational potential and
thus accumulates towards the centers of galaxies. Additionally, star
formation is proportional to the gas density thus producing increased
SF and enrichment towards the insides of the galaxy.

Environmental interactions are expected to modify the stellar
distributions, diluting any gradients, and possibly inducing
morphological transformations \citep{2012ApJ...751...61L}.
Additionally, recent simulations of the chemo-dynamical evolution of
dwarf galaxies have shown tidal interactions to be an efficient
mechanism to remove metallicity gradients \citep{2014A&A...564A.112N}.
Also, observations by \citet{2013ApJ...778..103H} have shown that
radial SFHs and metallicity gradients are
consistent with the SF in dwarf galaxies being quenched in the outer
regions due to limited gas availability.

Gradients in age and metallicity have  been seen for a wide range of
physical characteristics of dwarf galaxies.  The trend of younger
and higher metallicity stars increasing towards the centers of
galaxies spans many galactic properties such as total mass, gas mass,
metallicity, velocity dispersion and environment.  The fact that
gradients are consistently found even across a wide range of physical 
characteristics hints that population gradients are intrinsic to dwarf
galaxy formation and evolution.  However, at large radii, many of
these differences tend to disappear, suggesting some physical
processes may not affect all galactocentric radii equally (e.g.~the
UV background could affect the inner and outer regions differently).

\begin{deluxetable}{lccc}[H]
\tablefontsize{\scriptsize}
\tablecaption{Metallicity gradients}
\tablehead{
     \colhead{Galaxy}&
     \colhead{$d$[M/H]/$d\theta$}&
     \colhead{$d$[M/H]/$dr$}&
     \colhead{$d$[M/H]/$d(r/r_h$)}\\
	     \colhead{ name}& 
	     \colhead{(dex deg$^{-1}$)}&       
	     \colhead{(dex kpc$^{-1}$)}&  
	     \colhead{(dex per r$_h$})  } \\
 \startdata

Leo I     &-1.52	&-0.34 	&$-0.086 \pm    0.04$	\\
Leo II    &-11.55	&-3.02	&$-0.54 \pm    0.10 $	\\
IC 1613   &-2.48	&-0.20	&$-0.29 \pm    0.29 $	\\
Phoenix   &-1.84	&-0.26	&$-0.11 \pm    0.20 $
\enddata
\label{grad}
\end{deluxetable}

\begin{figure}[ht]
 \epsfig{file=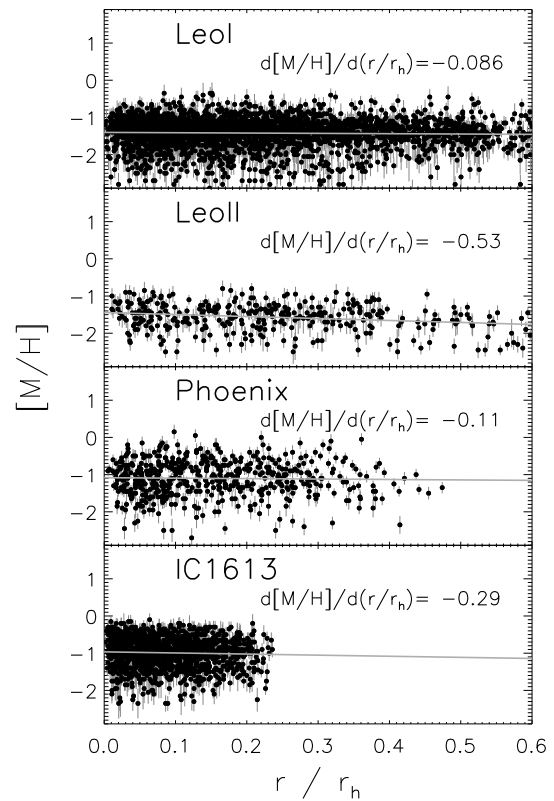, height=4.25in }
 \caption{The gradients are in units of dex of [M/H] per
   r$_h$, where r$_h$ is the 1/2 light radius as defined by \citet{2012AJ....144....4M}.   \label{gradients} } 
\end{figure}

The radial metallicity gradients we measured are reported in Table
\ref{grad}, and shown in Figure \ref{gradients}. We find metallicity
gradients in Leo I and Leo II. In Phoenix and IC 1613 the errors on
the gradients are consistent with no slope.  Much of the error is due
to the small radial extent the WFC3 field of view covers in these
galaxies, thus we cannot confirm or rule out a metallicity gradient in
Phoenix or IC 1613.

In Leo I age and metallicity gradients have been seen in various star
formation histories, variable star populations, and metallicity studies.  
\citet{2010MNRAS.404.1475H} found evidence for a radial gradient in
the 1-3 Gyr stellar population using near IR photometry.
\citet{2009A&A...500..735G} found a shallow metallicity gradient of
-0.27 dex kpc$^{-1}$ as part of their metallicity study of Leo I, which
is shallower than the gradient we measure, -0.34 dex kpc$^{-1}$, in
the same units.  In the work by \citet{2011ApJ...727...78K} gradients
for Leo I are reported as $d$[Fe/H]/$d$($r/r_c$) $=-0.11$ dex per
$r_c$, while we find a similar gradient for Leo I of $-0.09 \pm 0.07$
dex per $r_h$. Where the core radius ($r_c$=240 pc) or the half light
radius ($r_h$=251 pc) both give the same gradient value to 2
significant figures.  Our photometry covers Leo I out to $\sim$ 0.7 $r_h$,
while the work by \citet{2011ApJ...727...78K} extends $\sim$4 times
farther for Leo I. 

In Leo II, the wide field photometric survey by
\citet{2007AJ....134..835K} found evidence of a radial gradient in the 
HB morphology, where red HB stars are more centrally concentrated than
blue HB stars.  In the work by \citet{2011ApJ...727...78K} the
gradient for Leo II is reported as $d$[Fe/H]/$d$($r/r_c$) $-0.21$ dex
per $r_c$,  while we find a steeper gradient of $-0.54 \pm 0.10$ dex
per $r_h$.   Where the core radius ($r_c$=180 pc) and the 1/2 light
radius ($r_h$=176 pc) both give the same gradient value to 2
significant figures. Our photometry covers Leo II out to about 0.6 $r_h$,
while the work by \citet{2011ApJ...727...78K} extends $\sim$3 times
farther.   

In IC 1613 the wide field photometric study of \citet{2007AJ....134.1124B} 
looked at radial differences in the CMDs and concluded that there is
an age gradient, however, they could not confirm a metallicity
gradient in IC 1613 from their broadband ground based photometry. 

The radial SFH study of Phoenix by \citet{2009ApJ...705..704H,
2013ApJ...778..103H} measured the radial SF in Phoenix while
simultaneously deriving metallicities.  They found that Phoenix had
longer periods of SF in the central regions, and decreasing average
metallicities as function of radius, with the central regions having
an average [Fe/H] $\sim$ $-1.4$, decreasing outwards to $-1.7$, although
they do not quantify a metallicity gradient for the galaxy.

\section{Discussion}

\subsection{Morphological transitions and implications from the MDFs
  and chemical evolution models}

As the name suggests, a dwarf Transition galaxy is considered to be in
a state of morphological transition between a dIr and dSph type galaxy
\citep{1983ApJ...266L..21L,2001ApJ...559..754M,2003AJ....125.1926G,
2011ApJ...726...98K}.  This idea is supported by the presence of the 
morphology-density relationship, where dSph are mainly found in
dense environments close to a much larger primary while dIr and dTrans
are found in isolation \citep{2011ApJ...739....5W}. The dwarf galaxy
morphology segregation has led to an environment dependent
explanation invoking tidal stirring and ram pressure stripping to
transform a dIr close to a larger galaxy into a dTrans then  a dSph
over subsequent orbital passages around the primary
\citep{2007Natur.445..738M,2008ApJ...673..226P,2011ApJ...726...98K}.  

However, some dwarf galaxy properties conflict with the morphological
transition scenario.  DTrans have similar gas fractions
\citep{2011ApJ...743....8W}, and UV fluxes \citep{2011ApJS..192....6L}
as dIrs while the H$\alpha$ emission in dTrans is markedly
less and the typical stellar mass of dTrans is $\sim$ 3-4 times lower
than that of dIrs. DTrans are mostly located in isolated regions (like
dIrs), but dTrans luminosities, like dSphs, are lower than dIrs. This
led \citet{2011ApJ...743....8W} to suggest  that dTrans appear to be
lower mass versions of dIrs that lack recent SF, and not
an evolutionary bridge between dSphs and dIrs. Additionally, should
dTrans be progenitors of dSph  then they should be seen in similar
locations as both dSph and dIrs, but they are found only in the same
regions as dIrs.

Further evidence against a morphological transition theory comes from
\citet{2013ApJ...779..102K} who found systematic differences in the
MDF shapes of dIrs and dSphs. They found the dSph MDF shapes were
systematically narrower and more peaked while dIrs tend to have wider
MDFs that more closely resemble a simple chemical evolution model. The
narrower dSph MDF is best fit by chemical evolution models with large
amounts of accretion which is at odds with the proposed mechanism to
transform dIrs into dSphs, i.e.~gas stripping.  

Accretion has been shown to be an important component in dwarf galaxy
evolution \citep{2014MNRAS.443.3809B}.  Also, the analytical `Best
Accretion' model of \citet{1975VA.....19..299L} imply that dSph tend
to require large accretion parameters (M values) to fit the observed
MDFs \citep{2013ApJ...779..102K}. However, the morphology-density
relation points to gas stripping and tidal disruption as  important
factors in shaping dSph.  Simulations have also shown that gas
stripping and tidal disruption is integral in creating many of the the
dSph seen today \citep{2006MNRAS.369.1021M}.  Further examination of
the Best Accretion model of \citet{1975VA.....19..299L} shows that
there is no physical basis behind the form of Equation (3) other
than that it produces the desired MDF shapes.  Equation (3) simply
assumes a non-linear relation between gas mass and the stellar mass
that follows a quadratic relation that peaks then decays as a function
of stellar mass. Due to the contradictory implications from the accretion
model and the gas stripping caused by gravitational interactions known
to be integral in shaping dSphs, in addition to the lack of physical
meaning behind the functional form of the accretion model we suggest
that the interpretation of the accretion portion of the accretion model
should be left for more sophisticated chemical evolution models.  In
lieu of drawing interpretations from the models themselves, we can use the
differences in the model parameters in the context of the dynamic
histories of each galaxy to inform our discussion.

Examining the dSph type galaxies in our sample (Leo I and II), we find
that the MDF of Leo II is more similar to the MDFs from IC 1613 (dIr)
and Phoenix (dTrans) than to Leo I's. Leo II is also one of the more
distant dSph associated with the MW (233 kpc), although it is slightly
closer than Leo I (254 kpc) \citep{2012AJ....144....4M}. Dynamical
studies of Leo II \citep{2007AJ....134..566K} and wide field
photometric studies \citep{2007AJ....134.1938C} show no significant
signs of tidal distortion. \citet{2011ApJ...741..100L} studied Leo
II's proper motion and found nearly all of its velocity is in a
tangential component ($v_{tan}=265.2$ km s$^{-1}$) with only a small
radial component ($v_{rad}=21.5$ km s$^{-1}$). 
\citet{2011ApJ...741..100L} take this to mean that Leo II either has a
low-eccentricity orbit, or is near perigalacticon or apogalacticon,
and the lack of evidence for tidal disruptions
\citep{2007AJ....134..566K,2007AJ....134.1938C} likely rules out the latter.

Currently the only known mechanism to transform a rotating dIr into a
dSph is through gravitational influence from a larger primary
\citep{2001ApJ...559..754M}.  Consequently, it can be assumed that Leo
II has experienced some tidal interactions with the MW in the past.
However, Leo II shows no signs of dynamical interactions, beside the
fact that it is a dSph. Moreover the similarity of the MDF of Leo II, IC
1613 and Phoenix suggest these galaxies experienced comparable
conditions during their chemical evolution. Since IC 1613 and Phoenix
are isolated and evolved without outside influence, it could be
inferred that Leo II also chemically evolved without dynamical
influence from the MW.  The SFH of Leo II shows that 95$\%$ of the
stars were formed over 6.3 Gyr ago \citep{2014ApJ...789..147W}; with
so many of the stars formed at early times Leo II has had ample time
to be transformed from a disk to a spheroid after most of the chemical
evolution occurred. We do find the accretion + truncation model to be
the best fit for Leo II, where there the cutoff on the metal-rich
side of Leo II's MDF is shallower than predicted from the accretion model.
We interpret this as an indication of ram-pressure stripping; however,
in Leo II, the cutoff is not nearly as drastic as Leo I, indicating
that most of the chemical evolution had occurred before the MW began
stripping and truncating Leo II's SF. While this is by no means direct
proof, the evidence supports the idea that Leo II chemically evolved
without strong dynamical influence from the MW.

Leo I on the other hand shows  evidence of strong tidal interactions
with the MW. The proper motions and dynamic modeling of Leo I show it
passed into the MW's potential 2.33 Gyr ago, reaching its pericentric  
approach 1.05 Gyr ago \citep{2008ApJ...675..201M,2013ApJ...768..139S}.
Additionally,  the dynamic results correlate well with the CMD based
SFHs.  Leo I shows slow ancient SF, with an increase
to its highest rate 2-3 Gyr ago, after which the galaxy
stopped forming stars 1 Gyr ago \citep{1999AJ....118.2245G,
1999AJ....117.2199C,2000MNRAS.317..831H,2002ASPC..274..450D,
2014ApJ...789..147W}.  \citet{2013ApJ...768..139S} suggest that the
increased SF at 3 Gyr and the abrupt stop at 1 Gyr could have been
caused by ram pressure compression or gravitational torques exerted by
the MW as the galaxy passed into the MW's potential.  For Leo I, the
dynamical history has had a significant impact on the SFH, and by
extension on the chemical evolution.

We interpret the sharp difference in the M parameters along with the
dynamic information in the literature on Leo I, Leo II, IC 1613 and
Phoenix, as possible evidence that the galaxies chemically evolved
under different conditions.  The first is passive chemical evolution,
where gas rich, star forming galaxies (i.e.~dIr) gradually truncate
their star formation as the gas supply is exhausted, evolving into a
dTrans, producing a broad MDF in the process. This type of evolution
is what is most likely what is seen in Leo II (dSph), Phoenix (dTrans)
and IC 1613 (dIr), at various stages in the process.  Leo II formed
the majority of its stars over 6 Gyr ago and the MDF suggests it
chemically evolved without significant interactions before it was
transformed into a dSph by the MW. Phoenix appears to be in the
process of passively evolving; the dynamics and SFH show no indication
of interaction. Instead the SFH shows a slow decrease in SF up to a
few 100 Myr ago when SF stopped. Additionally, IC 1613 is isolated
from both the MW and M31, and does not show dynamical evidence of
interactions, and the MDF of IC 1613 is similar in shape to Leo II and
Phoenix.  It is known that dSph are created through the gravitational
influence of a larger primary, if the interaction occurs concurrent
with the chemical evolution it will be reflected in the MDF.  Previous
studies have shown that the MW's influence on Leo I is reflected in
the SFH and dynamics; we are suggesting it is also shown in the narrow
MDF.

\subsection{MDF evolution }

One understated assumption in analytic chemical evolution models is
that each model represents the end point of that galaxy's evolution.
However, dIr and dTrans are not at the end of their chemical evolution
since they are still forming stars and enriching the ISM.  Therefore
we must take care when comparing the MDFs of active galaxies such as
IC 1613 and Phoenix, to the dSphs like Leo I and II, which arguably,
have completed their chemical evolution 1 to 6 Gyr ago respectively. 
Additionally, for dSph the morphology-density relation suggest that
the removal of gas interrupted the chemical evolution altering the
chemistry to no longer reflect the end state of chemical evolution
models.

These various endpoints can be seen in our galaxy sample.  Leo I shows
a MDF truncated on the metal rich side indicative of a halted chemical
evolution. Leo II shows signs of mild truncation, as the CEM slightly
overpredicts the number of metal rich stars and the accretion +
truncation model was the best fit model for Leo II.   As we discussed
previously, Leo II is within the virial radius of the MW, but has a
mostly circular orbit at a large distance, therefore it has been
weakly affected by the MWs gravity, as the mild truncation shows.  

As a dIr, IC 1613 is still actively forming stars and enriching
its ISM, therefore we would not expect the MDF to reflect a fully
completed chemical evolution. The ongoing enrichment in IC 1613
manifests itself with fewer metal rich stars than predicted by the
CEM. 

Phoenix, on the other hand, is the only galaxy to nearly match the
metal rich end of the CEM.  Phoenix is a transition galaxy, that
expelled its last remaining gas reserves with the final burst of star
formation $\sim$ 100 Myr ago \citep{2007ApJ...659..331Y}. The isolated
location and apparent exhaustion of gas imply that Phoenix has
completed its chemical evolution. 

One other consideration is that our field of view is small for all of
the galaxies in our sample (covering 5-20$\%$ of the various
galaxies).  Given that many dwarf galaxies show metallicity gradients,
and variations in SFH as a function of radius, the MDFs will also
change as a function of radius, and the solutions to the CEMs cannot
be applied to the entire galaxy.

\section{Conclusion}
We measured the metallicity distribution functions of Leo I, Leo II,
IC 1613 and Phoenix dwarf galaxies by measuring individual stellar
metallicities, and by modeling the stellar population to create
synthetic MDFs for each galaxy.  We find the synthetic MDFs to be a
better representation of the overall metallicity distribution, because
this method cuts down the photometric scatter propagated into the
photometric metallicities.  

We fit each MDF with chemical evolution models.  We find the simple
model to be a poor fit for all of our dwarf galaxies, while the
accretion + truncation model is the best fit for the synthetic MDFs
for all four of the galaxies. The fact that all of our galaxies are
best fit by the accretion + truncation model reflects the fact that
most galaxies do not make it to the natural completion of the chemical
evolution, either because they are still actively forming stars (dIrs)
or because they were truncated by other processes such as ram pressure
stripping (dSphs).  

We find a similar accretion parameter, (M$\sim$4), for Leo II, IC 1613
and Phoenix despite the fact that they all have different masses, SFH,
morphologies and average metallicities.   We interpret the resemblance
in their MDFs as an indication that their chemical evolutions occurred
under similar conditions, which indicates that Leo II completed most
of its chemical evolution in isolation before it was significantly
tidally disrupted and transformed into a dSph type galaxy by the MW. 

Leo I has the narrowest MDF, a much larger accretion parameter, and
shows significant evidence for interactions with the MW at the same
time as the galaxy was forming stars and chemically evolving.  We
suggest that the MDFs can reveal dynamical interactions if they occur
in concert with SF and chemical evolution, or if the galaxy chemically
evolved in relative isolation. The narrower MDFs are indicative of
interactions shaping the galaxy's present morphology, while a broader
MDF indicates a passive evolution.  The differences in the MDFs could
be a way to distinguish between the two formation pathways. To further
this theory we would need to examine the MDFs and dynamic histories of
many LG dwarf galaxies.

Additionally, we measured metallicity gradients for Leo I and Leo II.
We see some evidence of gradients in Phoenix and IC 1613, however our
data does not cover the radial extent required to determine a
metallicity gradient greater than the error bars.

Support for program 12304 was provided by NASA through a grant from
the Space Telescope Science Institute, which is operated by the
Association of Universities for Research in Astronomy, Inc., under
NASA contract NAS 5-26555.


\begin{thebibliography}{82}
\expandafter\ifx\csname natexlab\endcsname\relax\def\natexlab#1{#1}\fi

\bibitem[{{Battaglia} {et~al.}(2012{\natexlab{a}}){Battaglia}, {Irwin},
  {Tolstoy}, {de Boer}, \& {Mateo}}]{2012ApJ...761L..31B}
{Battaglia}, G., {Irwin}, M., {Tolstoy}, E., {de Boer}, T., \& {Mateo}, M.
  2012{\natexlab{a}}, \apjl, 761, L31

\bibitem[{{Battaglia} {et~al.}(2012{\natexlab{b}}){Battaglia}, {Rejkuba},
  {Tolstoy}, {Irwin}, \& {Beccari}}]{2012MNRAS.424.1113B}
{Battaglia}, G., {Rejkuba}, M., {Tolstoy}, E., {Irwin}, M.~J., \& {Beccari}, G.
  2012{\natexlab{b}}, \mnras, 424, 1113

\bibitem[{{Battaglia} {et~al.}(2011){Battaglia}, {Tolstoy}, {Helmi}, {Irwin},
  {Parisi}, {Hill}, \& {Jablonka}}]{2011MNRAS.411.1013B}
{Battaglia}, G., {Tolstoy}, E., {Helmi}, A., {et~al.} 2011, \mnras, 411, 1013

\bibitem[{{Battaglia} {et~al.}(2006){Battaglia}, {Tolstoy}, {Helmi}, {Irwin},
  {Letarte}, {Jablonka}, {Hill}, {Venn}, {Shetrone}, {Arimoto}, {Primas},
  {Kaufer}, {Francois}, {Szeifert}, {Abel}, \&
  {Sadakane}}]{2006A&A...459..423B}
---. 2006, \aap, 459, 423

\bibitem[{{Bernard} {et~al.}(2007){Bernard}, {Aparicio}, {Gallart},
  {Padilla-Torres}, \& {Panniello}}]{2007AJ....134.1124B}
{Bernard}, E.~J., {Aparicio}, A., {Gallart}, C., {Padilla-Torres}, C.~P., \&
  {Panniello}, M. 2007, \aj, 134, 1124

\bibitem[{{Bianchi} {et~al.}(2012){Bianchi}, {Efremova}, {Hodge}, {Massey}, \&
  {Olsen}}]{2012AJ....143...74B}
{Bianchi}, L., {Efremova}, B., {Hodge}, P., {Massey}, P., \& {Olsen}, K.~A.~G.
  2012, \aj, 143, 74

\bibitem[{{Bosler} {et~al.}(2007){Bosler}, {Smecker-Hane}, \&
  {Stetson}}]{2007MNRAS.378..318B}
{Bosler}, T.~L., {Smecker-Hane}, T.~A., \& {Stetson}, P.~B. 2007, \mnras, 378,
  318

\bibitem[{{Brook} {et~al.}(2014){Brook}, {Stinson}, {Gibson}, {Shen},
  {Macci{\`o}}, {Obreja}, {Wadsley}, \& {Quinn}}]{2014MNRAS.443.3809B}
{Brook}, C.~B., {Stinson}, G., {Gibson}, B.~K., {et~al.} 2014, \mnras, 443,
  3809

\bibitem[{{Caputo} {et~al.}(1999){Caputo}, {Cassisi}, {Castellani}, {Marconi},
  \& {Santolamazza}}]{1999AJ....117.2199C}
{Caputo}, F., {Cassisi}, S., {Castellani}, M., {Marconi}, G., \&
  {Santolamazza}, P. 1999, \aj, 117, 2199

\bibitem[{{Coleman} {et~al.}(2007){Coleman}, {Jordi}, {Rix}, {Grebel}, \&
  {Koch}}]{2007AJ....134.1938C}
{Coleman}, M.~G., {Jordi}, K., {Rix}, H.-W., {Grebel}, E.~K., \& {Koch}, A.
  2007, \aj, 134, 1938

\bibitem[{{Dolphin}(2000)}]{2000PASP..112.1383D}
{Dolphin}, A.~E. 2000, \pasp, 112, 1383

\bibitem[{{Dolphin}(2002)}]{2002ASPC..274..450D}
{Dolphin}, A.~E. 2002, in Astronomical Society of the Pacific Conference
  Series, Vol. 274, Observed HR Diagrams and Stellar Evolution, ed.
  T.~{Lejeune} \& J.~{Fernandes}, 450

\bibitem[{{Dotter} {et~al.}(2008){Dotter}, {Chaboyer}, {Jevremovi{\'c}},
  {Kostov}, {Baron}, \& {Ferguson}}]{2008ApJS..178...89D}
{Dotter}, A., {Chaboyer}, B., {Jevremovi{\'c}}, D., {et~al.} 2008, \apjs, 178,
  89

\bibitem[{{Gallart} {et~al.}(1999){Gallart}, {Freedman}, {Aparicio},
  {Bertelli}, \& {Chiosi}}]{1999AJ....118.2245G}
{Gallart}, C., {Freedman}, W.~L., {Aparicio}, A., {Bertelli}, G., \& {Chiosi},
  C. 1999, \aj, 118, 2245

\bibitem[{{Garcia} {et~al.}(2010){Garcia}, {Herrero}, {Castro}, {Corral}, \&
  {Rosenberg}}]{2010A&A...523A..23G}
{Garcia}, M., {Herrero}, A., {Castro}, N., {Corral}, L., \& {Rosenberg}, A.
  2010, \aap, 523, A23

\bibitem[{{Grcevich} \& {Putman}(2009)}]{2009ApJ...696..385G}
{Grcevich}, J., \& {Putman}, M.~E. 2009, \apj, 696, 385

\bibitem[{{Grebel} {et~al.}(2003){Grebel}, {Gallagher}, \&
  {Harbeck}}]{2003AJ....125.1926G}
{Grebel}, E.~K., {Gallagher}, III, J.~S., \& {Harbeck}, D. 2003, \aj, 125, 1926

\bibitem[{{Gullieuszik} {et~al.}(2008){Gullieuszik}, {Held}, {Rizzi},
  {Girardi}, {Marigo}, \& {Momany}}]{2008MNRAS.388.1185G}
{Gullieuszik}, M., {Held}, E.~V., {Rizzi}, L., {et~al.} 2008, \mnras, 388, 1185

\bibitem[{{Gullieuszik} {et~al.}(2009){Gullieuszik}, {Held}, {Saviane}, \&
  {Rizzi}}]{2009A&A...500..735G}
{Gullieuszik}, M., {Held}, E.~V., {Saviane}, I., \& {Rizzi}, L. 2009, \aap,
  500, 735

\bibitem[{{Harbeck} {et~al.}(2001){Harbeck}, {Grebel}, {Holtzman},
  {Guhathakurta}, {Brandner}, {Geisler}, {Sarajedini}, {Dolphin},
  {Hurley-Keller}, \& {Mateo}}]{2001AJ....122.3092H}
{Harbeck}, D., {Grebel}, E.~K., {Holtzman}, J., {et~al.} 2001, \aj, 122, 3092

\bibitem[{{Harrington} \& {Wilson}(1950)}]{1950PASP...62..118H}
{Harrington}, R.~G., \& {Wilson}, A.~G. 1950, \pasp, 62, 118

\bibitem[{{Held} {et~al.}(2010){Held}, {Gullieuszik}, {Rizzi}, {Girardi},
  {Marigo}, \& {Saviane}}]{2010MNRAS.404.1475H}
{Held}, E.~V., {Gullieuszik}, M., {Rizzi}, L., {et~al.} 2010, \mnras, 404, 1475

\bibitem[{{Held} {et~al.}(1999){Held}, {Saviane}, \&
  {Momany}}]{1999A&A...345..747H}
{Held}, E.~V., {Saviane}, I., \& {Momany}, Y. 1999, \aap, 345, 747

\bibitem[{{Hendricks} {et~al.}(2014){Hendricks}, {Koch}, {Lanfranchi},
  {Boeche}, {Walker}, {Johnson}, {Pe{\~n}arrubia}, \&
  {Gilmore}}]{2014ApJ...785..102H}
{Hendricks}, B., {Koch}, A., {Lanfranchi}, G.~A., {et~al.} 2014, \apj, 785, 102

\bibitem[{{Hernandez} {et~al.}(2000){Hernandez}, {Gilmore}, \&
  {Valls-Gabaud}}]{2000MNRAS.317..831H}
{Hernandez}, X., {Gilmore}, G., \& {Valls-Gabaud}, D. 2000, \mnras, 317, 831

\bibitem[{{Hidalgo} {et~al.}(2009){Hidalgo}, {Aparicio},
  {Mart{\'{\i}}nez-Delgado}, \& {Gallart}}]{2009ApJ...705..704H}
{Hidalgo}, S.~L., {Aparicio}, A., {Mart{\'{\i}}nez-Delgado}, D., \& {Gallart},
  C. 2009, \apj, 705, 704

\bibitem[{{Hidalgo} {et~al.}(2013){Hidalgo}, {Monelli}, {Aparicio}, {Gallart},
  {Skillman}, {Cassisi}, {Bernard}, {Mayer}, {Stetson}, {Cole}, \&
  {Dolphin}}]{2013ApJ...778..103H}
{Hidalgo}, S.~L., {Monelli}, M., {Aparicio}, A., {et~al.} 2013, \apj, 778, 103

\bibitem[{{Holtzman}(2008)}]{2008hst..prop11729H}
{Holtzman}, J. 2008, in HST Proposal, 11729

\bibitem[{{Holtzman} {et~al.}(2000){Holtzman}, {Smith}, \&
  {Grillmair}}]{2000AJ....120.3060H}
{Holtzman}, J.~A., {Smith}, G.~H., \& {Grillmair}, C. 2000, \aj, 120, 3060

\bibitem[{{Kazantzidis} {et~al.}(2011){Kazantzidis}, {{\L}okas}, {Callegari},
  {Mayer}, \& {Moustakas}}]{2011ApJ...726...98K}
{Kazantzidis}, S., {{\L}okas}, E.~L., {Callegari}, S., {Mayer}, L., \&
  {Moustakas}, L.~A. 2011, \apj, 726, 98

\bibitem[{{Kirby} {et~al.}(2013){Kirby}, {Cohen}, {Guhathakurta}, {Cheng},
  {Bullock}, \& {Gallazzi}}]{2013ApJ...779..102K}
{Kirby}, E.~N., {Cohen}, J.~G., {Guhathakurta}, P., {et~al.} 2013, \apj, 779,
  102

\bibitem[{{Kirby} {et~al.}(2011{\natexlab{a}}){Kirby}, {Cohen}, {Smith},
  {Majewski}, {Sohn}, \& {Guhathakurta}}]{2011ApJ...727...79K}
{Kirby}, E.~N., {Cohen}, J.~G., {Smith}, G.~H., {et~al.} 2011{\natexlab{a}},
  \apj, 727, 79

\bibitem[{{Kirby} {et~al.}(2011{\natexlab{b}}){Kirby}, {Lanfranchi}, {Simon},
  {Cohen}, \& {Guhathakurta}}]{2011ApJ...727...78K}
{Kirby}, E.~N., {Lanfranchi}, G.~A., {Simon}, J.~D., {Cohen}, J.~G., \&
  {Guhathakurta}, P. 2011{\natexlab{b}}, \apj, 727, 78

\bibitem[{{Koch} {et~al.}(2013){Koch}, {Feltzing}, {Ad{\'e}n}, \&
  {Matteucci}}]{2013A&A...554A...5K}
{Koch}, A., {Feltzing}, S., {Ad{\'e}n}, D., \& {Matteucci}, F. 2013, \aap, 554,
  A5

\bibitem[{{Koch} {et~al.}(2007{\natexlab{a}}){Koch}, {Grebel}, {Kleyna},
  {Wilkinson}, {Harbeck}, {Gilmore}, {Wyse}, \& {Evans}}]{2007AJ....133..270K}
{Koch}, A., {Grebel}, E.~K., {Kleyna}, J.~T., {et~al.} 2007{\natexlab{a}}, \aj,
  133, 270

\bibitem[{{Koch} {et~al.}(2007{\natexlab{b}}){Koch}, {Kleyna}, {Wilkinson},
  {Grebel}, {Gilmore}, {Evans}, {Wyse}, \& {Harbeck}}]{2007AJ....134..566K}
{Koch}, A., {Kleyna}, J.~T., {Wilkinson}, M.~I., {et~al.} 2007{\natexlab{b}},
  \aj, 134, 566

\bibitem[{{Komiyama} {et~al.}(2007){Komiyama}, {Doi}, {Furusawa}, {Hamabe},
  {Imi}, {Kimura}, {Miyazaki}, {Nakata}, {Okada}, {Okamura}, {Ouchi},
  {Sekiguchi}, {Shimasaku}, {Yagi}, \& {Yasuda}}]{2007AJ....134..835K}
{Komiyama}, Y., {Doi}, M., {Furusawa}, H., {et~al.} 2007, \aj, 134, 835

\bibitem[{{Kroupa}(2001)}]{2001MNRAS.322..231K}
{Kroupa}, P. 2001, \mnras, 322, 231

\bibitem[{{Lake} \& {Skillman}(1989)}]{1989AJ.....98.1274L}
{Lake}, G., \& {Skillman}, E.~D. 1989, \aj, 98, 1274

\bibitem[{{Lanfranchi} \& {Matteucci}(2010)}]{2010A&A...512A..85L}
{Lanfranchi}, G.~A., \& {Matteucci}, F. 2010, \aap, 512, A85

\bibitem[{{Lee} {et~al.}(2006){Lee}, {Skillman}, {Cannon}, {Jackson}, {Gehrz},
  {Polomski}, \& {Woodward}}]{2006ApJ...647..970L}
{Lee}, H., {Skillman}, E.~D., {Cannon}, J.~M., {et~al.} 2006, \apj, 647, 970

\bibitem[{{Lee} {et~al.}(2011){Lee}, {Gil de Paz}, {Kennicutt}, {Bothwell},
  {Dalcanton}, {Jos{\'e} G.~Funes S.}, {Johnson}, {Sakai}, {Skillman},
  {Tremonti}, \& {van Zee}}]{2011ApJS..192....6L}
{Lee}, J.~C., {Gil de Paz}, A., {Kennicutt}, Jr., R.~C., {et~al.} 2011, \apjs,
  192, 6

\bibitem[{{Lemasle} {et~al.}(2014){Lemasle}, {de Boer}, {Hill}, {Tolstoy},
  {Irwin}, {Jablonka}, {Venn}, {Battaglia}, {Starkenburg}, {Shetrone},
  {Letarte}, {Fran{\c c}ois}, {Helmi}, {Primas}, {Kaufer}, \&
  {Szeifert}}]{2014A&A...572A..88L}
{Lemasle}, B., {de Boer}, T.~J.~L., {Hill}, V., {et~al.} 2014, \aap, 572, A88

\bibitem[{{L{\'e}pine} {et~al.}(2011){L{\'e}pine}, {Koch}, {Rich}, \&
  {Kuijken}}]{2011ApJ...741..100L}
{L{\'e}pine}, S., {Koch}, A., {Rich}, R.~M., \& {Kuijken}, K. 2011, \apj, 741,
  100

\bibitem[{{Lin} \& {Faber}(1983)}]{1983ApJ...266L..21L}
{Lin}, D.~N.~C., \& {Faber}, S.~M. 1983, \apjl, 266, L21

\bibitem[{{{\L}okas} {et~al.}(2012){{\L}okas}, {Majewski}, {Kazantzidis},
  {Mayer}, {Carlin}, {Nidever}, \& {Moustakas}}]{2012ApJ...751...61L}
{{\L}okas}, E.~L., {Majewski}, S.~R., {Kazantzidis}, S., {et~al.} 2012, \apj,
  751, 61

\bibitem[{{Lynden-Bell}(1975)}]{1975VA.....19..299L}
{Lynden-Bell}, D. 1975, Vistas in Astronomy, 19, 299

\bibitem[{{Marcolini} {et~al.}(2008){Marcolini}, {D'Ercole}, {Battaglia}, \&
  {Gibson}}]{2008MNRAS.386.2173M}
{Marcolini}, A., {D'Ercole}, A., {Battaglia}, G., \& {Gibson}, B.~K. 2008,
  \mnras, 386, 2173

\bibitem[{{Marcolini} {et~al.}(2006){Marcolini}, {D'Ercole}, {Brighenti}, \&
  {Recchi}}]{2006MNRAS.371..643M}
{Marcolini}, A., {D'Ercole}, A., {Brighenti}, F., \& {Recchi}, S. 2006, \mnras,
  371, 643

\bibitem[{{Mart{\'{\i}}nez-Delgado} {et~al.}(1999){Mart{\'{\i}}nez-Delgado},
  {Gallart}, \& {Aparicio}}]{1999AJ....118..862M}
{Mart{\'{\i}}nez-Delgado}, D., {Gallart}, C., \& {Aparicio}, A. 1999, \aj, 118,
  862

\bibitem[{{Mateo} {et~al.}(2008){Mateo}, {Olszewski}, \&
  {Walker}}]{2008ApJ...675..201M}
{Mateo}, M., {Olszewski}, E.~W., \& {Walker}, M.~G. 2008, \apj, 675, 201

\bibitem[{{Mateo}(1998)}]{1998ARA&A..36..435M}
{Mateo}, M.~L. 1998, \araa, 36, 435

\bibitem[{{Mayer} {et~al.}(2001){Mayer}, {Governato}, {Colpi}, {Moore},
  {Quinn}, {Wadsley}, {Stadel}, \& {Lake}}]{2001ApJ...559..754M}
{Mayer}, L., {Governato}, F., {Colpi}, M., {et~al.} 2001, \apj, 559, 754

\bibitem[{{Mayer} {et~al.}(2007){Mayer}, {Kazantzidis}, {Mastropietro}, \&
  {Wadsley}}]{2007Natur.445..738M}
{Mayer}, L., {Kazantzidis}, S., {Mastropietro}, C., \& {Wadsley}, J. 2007,
  \nat, 445, 738

\bibitem[{{Mayer} {et~al.}(2006){Mayer}, {Mastropietro}, {Wadsley}, {Stadel},
  \& {Moore}}]{2006MNRAS.369.1021M}
{Mayer}, L., {Mastropietro}, C., {Wadsley}, J., {Stadel}, J., \& {Moore}, B.
  2006, \mnras, 369, 1021

\bibitem[{{McConnachie}(2012)}]{2012AJ....144....4M}
{McConnachie}, A.~W. 2012, \aj, 144, 4

\bibitem[{{Monelli} {et~al.}(2012){Monelli}, {Bernard}, {Gallart},
  {Fiorentino}, {Drozdovsky}, {Aparicio}, {Bono}, {Cassisi}, {Skillman}, \&
  {Stetson}}]{2012MNRAS.422...89M}
{Monelli}, M., {Bernard}, E.~J., {Gallart}, C., {et~al.} 2012, \mnras, 422, 89

\bibitem[{{Nichols} {et~al.}(2014){Nichols}, {Revaz}, \&
  {Jablonka}}]{2014A&A...564A.112N}
{Nichols}, M., {Revaz}, Y., \& {Jablonka}, P. 2014, \aap, 564, A112

\bibitem[{{Ordo{\~n}ez} {et~al.}(2014){Ordo{\~n}ez}, {Yang}, \&
  {Sarajedini}}]{2014ApJ...786..147O}
{Ordo{\~n}ez}, A.~J., {Yang}, S.-C., \& {Sarajedini}, A. 2014, \apj, 786, 147

\bibitem[{{Ortolani} \& {Gratton}(1988)}]{1988PASP..100.1405O}
{Ortolani}, S., \& {Gratton}, R.~G. 1988, \pasp, 100, 1405

\bibitem[{{Pagel}(1997)}]{1997nceg.book.....P}
{Pagel}, B.~E.~J. 1997, {Nucleosynthesis and Chemical Evolution of Galaxies}

\bibitem[{{Pe{\~n}arrubia} {et~al.}(2008){Pe{\~n}arrubia}, {Navarro}, \&
  {McConnachie}}]{2008ApJ...673..226P}
{Pe{\~n}arrubia}, J., {Navarro}, J.~F., \& {McConnachie}, A.~W. 2008, \apj,
  673, 226

\bibitem[{{Prantzos}(2008)}]{2008A&A...489..525P}
{Prantzos}, N. 2008, \aap, 489, 525

\bibitem[{{Revaz} {et~al.}(2009){Revaz}, {Jablonka}, {Sawala}, {Hill},
  {Letarte}, {Irwin}, {Battaglia}, {Helmi}, {Shetrone}, {Tolstoy}, \&
  {Venn}}]{2009A&A...501..189R}
{Revaz}, Y., {Jablonka}, P., {Sawala}, T., {et~al.} 2009, \aap, 501, 189

\bibitem[{{Ross} {et~al.}(2014){Ross}, {Holtzman}, {Anthony-Twarog}, {Bond},
  {Twarog}, {Saha}, \& {Walker}}]{2014AJ....147....4R}
{Ross}, T.~L., {Holtzman}, J.~A., {Anthony-Twarog}, B.~J., {et~al.} 2014, \aj,
  147, 4

\bibitem[{{Shetrone} {et~al.}(2003){Shetrone}, {Venn}, {Tolstoy}, {Primas},
  {Hill}, \& {Kaufer}}]{2003AJ....125..684S}
{Shetrone}, M., {Venn}, K.~A., {Tolstoy}, E., {et~al.} 2003, \aj, 125, 684

\bibitem[{{Shetrone} {et~al.}(2001){Shetrone}, {C{\^o}t{\'e}}, \&
  {Sargent}}]{2001ApJ...548..592S}
{Shetrone}, M.~D., {C{\^o}t{\'e}}, P., \& {Sargent}, W.~L.~W. 2001, \apj, 548,
  592

\bibitem[{{Shetrone} {et~al.}(2009){Shetrone}, {Siegel}, {Cook}, \&
  {Bosler}}]{2009AJ....137...62S}
{Shetrone}, M.~D., {Siegel}, M.~H., {Cook}, D.~O., \& {Bosler}, T. 2009, \aj,
  137, 62

\bibitem[{{Silich} {et~al.}(2006){Silich}, {Lozinskaya}, {Moiseev},
  {Podorvanuk}, {Rosado}, {Borissova}, \&
  {Valdez-Gutierrez}}]{2006A&A...448..123S}
{Silich}, S., {Lozinskaya}, T., {Moiseev}, A., {et~al.} 2006, \aap, 448, 123

\bibitem[{{Skillman} {et~al.}(2003){Skillman}, {Tolstoy}, {Cole}, {Dolphin},
  {Saha}, {Gallagher}, {Dohm-Palmer}, \& {Mateo}}]{2003ApJ...596..253S}
{Skillman}, E.~D., {Tolstoy}, E., {Cole}, A.~A., {et~al.} 2003, \apj, 596, 253

\bibitem[{{Skillman} {et~al.}(2014){Skillman}, {Hidalgo}, {Weisz}, {Monelli},
  {Gallart}, {Aparicio}, {Bernard}, {Boylan-Kolchin}, {Cassisi}, {Cole},
  {Dolphin}, {Ferguson}, {Mayer}, {Navarro}, {Stetson}, \&
  {Tolstoy}}]{2014ApJ...786...44S}
{Skillman}, E.~D., {Hidalgo}, S.~L., {Weisz}, D.~R., {et~al.} 2014, \apj, 786,
  44

\bibitem[{{Sohn} {et~al.}(2013){Sohn}, {Besla}, {van der Marel},
  {Boylan-Kolchin}, {Majewski}, \& {Bullock}}]{2013ApJ...768..139S}
{Sohn}, S.~T., {Besla}, G., {van der Marel}, R.~P., {et~al.} 2013, \apj, 768,
  139

\bibitem[{{Spekkens} {et~al.}(2014){Spekkens}, {Urbancic}, {Mason}, {Willman},
  \& {Aguirre}}]{2014ApJ...795L...5S}
{Spekkens}, K., {Urbancic}, N., {Mason}, B.~S., {Willman}, B., \& {Aguirre},
  J.~E. 2014, \apjl, 795, L5

\bibitem[{{Starkenburg} {et~al.}(2013){Starkenburg}, {Hill}, {Tolstoy},
  {Fran{\c c}ois}, {Irwin}, {Boschman}, {Venn}, {de Boer}, {Lemasle},
  {Jablonka}, {Battaglia}, {Groot}, \& {Kaper}}]{2013A&A...549A..88S}
{Starkenburg}, E., {Hill}, V., {Tolstoy}, E., {et~al.} 2013, \aap, 549, A88

\bibitem[{{Tautvai{\v s}ien{\.e}} {et~al.}(2007){Tautvai{\v s}ien{\.e}},
  {Geisler}, {Wallerstein}, {Borissova}, {Bizyaev}, {Pagel}, {Charbonnel}, \&
  {Smith}}]{2007AJ....134.2318T}
{Tautvai{\v s}ien{\.e}}, G., {Geisler}, D., {Wallerstein}, G., {et~al.} 2007,
  \aj, 134, 2318

\bibitem[{{Tolstoy} {et~al.}(2009){Tolstoy}, {Hill}, \&
  {Tosi}}]{2009ARA&A..47..371T}
{Tolstoy}, E., {Hill}, V., \& {Tosi}, M. 2009, \araa, 47, 371

\bibitem[{{Tolstoy} {et~al.}(2003){Tolstoy}, {Venn}, {Shetrone}, {Primas},
  {Hill}, {Kaufer}, \& {Szeifert}}]{2003AJ....125..707T}
{Tolstoy}, E., {Venn}, K.~A., {Shetrone}, M., {et~al.} 2003, \aj, 125, 707

\bibitem[{{Tolstoy} {et~al.}(2004){Tolstoy}, {Irwin}, {Helmi}, {Battaglia},
  {Jablonka}, {Hill}, {Venn}, {Shetrone}, {Letarte}, {Cole}, {Primas},
  {Francois}, {Arimoto}, {Sadakane}, {Kaufer}, {Szeifert}, \&
  {Abel}}]{2004ApJ...617L.119T}
{Tolstoy}, E., {Irwin}, M.~J., {Helmi}, A., {et~al.} 2004, \apjl, 617, L119

\bibitem[{{Weisz} {et~al.}(2014){Weisz}, {Dolphin}, {Skillman}, {Holtzman},
  {Gilbert}, {Dalcanton}, \& {Williams}}]{2014ApJ...789..147W}
{Weisz}, D.~R., {Dolphin}, A.~E., {Skillman}, E.~D., {et~al.} 2014, \apj, 789,
  147

\bibitem[{{Weisz} {et~al.}(2011{\natexlab{a}}){Weisz}, {Dolphin}, {Dalcanton},
  {Skillman}, {Holtzman}, {Williams}, {Gilbert}, {Seth}, {Cole}, {Gogarten},
  {Rosema}, {Karachentsev}, {McQuinn}, \& {Zaritsky}}]{2011ApJ...743....8W}
{Weisz}, D.~R., {Dolphin}, A.~E., {Dalcanton}, J.~J., {et~al.}
  2011{\natexlab{a}}, \apj, 743, 8

\bibitem[{{Weisz} {et~al.}(2011{\natexlab{b}}){Weisz}, {Dalcanton}, {Williams},
  {Gilbert}, {Skillman}, {Seth}, {Dolphin}, {McQuinn}, {Gogarten}, {Holtzman},
  {Rosema}, {Cole}, {Karachentsev}, \& {Zaritsky}}]{2011ApJ...739....5W}
{Weisz}, D.~R., {Dalcanton}, J.~J., {Williams}, B.~F., {et~al.}
  2011{\natexlab{b}}, \apj, 739, 5

\bibitem[{{Young} {et~al.}(2007){Young}, {Skillman}, {Weisz}, \&
  {Dolphin}}]{2007ApJ...659..331Y}
{Young}, L.~M., {Skillman}, E.~D., {Weisz}, D.~R., \& {Dolphin}, A.~E. 2007,
  \apj, 659, 331

\end{thebibliography}

\end{document}